\documentclass[prd,preprint,eqsecnum,aps,amsfonts,amsmath,amssymb]{revtex4}
\allowdisplaybreaks
\begin{document}
\title{Energy losses by gravitational radiation in inspiralling \\
compact binaries to five halves post-Newtonian order}
\author{Luc Blanchet}
\affiliation{D\'epartement d'Astrophysique Relativiste et de Cosmologie,\\
Centre National de la Recherche Scientifique (UPR 176),\\
Observatoire de Paris, 92195 Meudon Cedex, France}
\date{\today}  
	
\begin{abstract}
This paper derives the total power or energy loss rate generated in the
form of gravitational waves by an inspiralling compact binary system to the
five halves post-Newtonian (2.5PN) approximation of general relativity.
Extending a recently developed gravitational-wave generation formalism
valid for arbitrary (slowly-moving) systems, we compute the mass
multipole moments of the system and the relevant tails present in the wave
zone to 2.5PN order.  In the case of two point-masses moving on a
quasi-circular orbit, we find that the 2.5PN contribution in the energy
loss rate is entirely due to tails. Relying on an energy balance argument
we derive the laws of variation of the instantaneous frequency and phase
of the binary. The 2.5PN order in the accumulated phase is significantly
large, being grossly of the same order of magnitude as the previous 2PN
order, but opposite in sign.  However finite mass effects at 2.5PN order
are small.  The results of this paper should be useful when analyzing
the data from inspiralling compact binaries in future gravitational-wave
detectors like VIRGO and LIGO.
\end{abstract}

\maketitle

\section{Introduction}
\label{sec:1}

Compact binaries in their late stage of evolution are very relativistic
systems in which the two compact objects (neutron stars or black holes)
orbit around eachother with velocities as large as 30~\% of that of light.
The gravitational radiation these systems emit during the inspiral phase
preceding the coalescence of the two objects is expected to be routinely
analyzed in future detectors like LIGO and VIRGO (see \cite{Th300,Sc89',Th94}
for reviews). Hundreds to tens of thousands of gravitational-wave cycles
should be monitored in the detectors' sensitive frequency bandwidth. The
combination of high orbital velocities and a large number of observed
rotations, together with the fact that the emitted waves are highly
predictable, implies that high-order relativistic (or post-Newtonian)
effects should show up in the gravitational signals observed by VIRGO
and LIGO \cite{KSc87,K89,LW90,3mn,FCh93,CF94,BSat95}. Alternatively, this
means that high-order post-Newtonian effects should be known in advance
so that they can be included in the construction of theoretical filters
(templates) to be cross correlated with the outputs of the detectors.

The relevant model for describing most of the observed inspiral phase is
a model of two point masses moving on a circular orbit.
Radiation reaction forces tend to circularize very rapidly the orbit. On
the other hand, point masses can be used in the case of non-rotating and
(initially) spherically symmetric compact objects up to a very high precision
\cite{D83a}.  This is due to a property owned by general relativity of
``effacing'' of the internal structure.  Even in the case of stars with
intrinsic rotations the dynamics of the binary is likely to be dominated by
post-Newtonian gravitational effects \cite{BDIWWi95}.

High-order post-Newtonian effects that are measurable are mainly those
affecting the orbital phase evolution of the binary, which in turn is
determined using a standard energy balance argument by the total power
emitted in the form of gravitational waves by the system at infinity, or
total luminosity in the waves. To what level in a post-Newtonian expansion
should we know the gravitational luminosity (or energy loss) in order to
guarantee an optimal detection of the signal (given some power spectral
density of the noise in a detector) is still unclear, but the theory of
black-hole perturbations can be used to gain insights in this problem.
Black-hole perturbations, which deal with the special case of a test mass
orbiting a massive black hole, have been recently the focus of intense activity
\cite{P93,CFPS93,TNaka94,Sasa94,TSasa94}. It emerges from this field that
neglecting even a so high approximation as the third post-Newtonian (3PN) one,
i.e. neglecting the relativistic corrections in the luminosity which are of
relative order $c^{-6}$ (or below), is likely to yield unacceptable systematic
errors in the data analysis of binary signals \cite{CFPS93,P95,CF95}.
This shows how relativistic are inspiralling compact binaries, as compared
for instance to the binary pulsar for which the Newtonian approximation in
the luminosity (Einstein quadrupole formula) is adequate. The post-Newtonian
theory is presently completed through the second post-Newtonian (2PN)
approximation, i.e., through relative order $c^{-4}$ (both in the waveform
and in the associated energy loss).  Two computations were performed to
this order, one by Blanchet, Damour and Iyer \cite{B95,BDI95} based on a
post-Minkowski matching formalism, and one by Will and Wiseman \cite{WWi95}
using an approach initiated by Epstein and Wagoner \cite{EW75} and generalized
by Thorne \cite{Th80}. The common result of these two computations for
the energy loss was summarized in Ref.~\cite{BDIWWi95}, and the waveform
can be found in \cite{BIWW95}.

In the present paper we develop the post-Minkowski matching formalism one
step beyond the work of Refs.~\cite{B95,BDI95} by computing the 2.5PN order
in the energy loss of an inspiralling binary. This entails extending both
Ref.~\cite{B95} on the general formalism valid for an arbitrary (slowly-moving)
source, and Ref.~\cite{BDI95} dealing with the specific application to the
binary. The computation of the waveform of the binary to 2.5PN order (the
square of which should give back the 2.5PN energy loss) will be left for
future work.

The post-Minkowski matching formalism is a wave generation formalism
which is especially suited for ``semi relativistic'' sources whose internal
velocities can reach 0.3c at most (say), as in the case of inspiralling compact
binaries (see \cite{Bhouches} for a review). The formalism combines (i)
an analytic post-Minkowskian approximation scheme for the computation of
the gravitational field in the exterior of the source where multipole
expansions can be used to simplify the problem, (ii) a direct post-Newtonian
approximation scheme for the resolution of the field equations inside the
near-zone of the source, and (iii) an asymptotic matching between both types
of solutions which is performed in the exterior part of the near zone. The
necessicity of using a post-Minkowskian approximation scheme first is because
its validity extends up to the regions far away from the system where the
observer is located, contrarily to the post-Newtonian approximation whose
validity is limited to the near zone. The exterior field is computed using an
algorithm developed in Ref.~\cite{BD86} which set on a general footing previous
investigations by Thorne~\cite{Th80} and Bonnor~\cite{Bo59}. The
implementation of the wave generation formalism (i)-(iii) was done at
first with 1PN accuracy in Ref.~\cite{BD89} which obtained the 1PN correction
terms in the mass-type quadrupole moment of the source (and in fact in all
the mass-type multipole moments). The dominant nonlinear contribution in the
radiation field was added in Ref.~\cite{BD92} and shown to be due to the
contribution in the wave zone of the well-known ``tail" effect. The inclusion
of this nonlinear contribution pushed the accuracy of the formalism to
1.5PN order in the energy loss.  The 2PN precision both in the energy
loss and waveform was reached in Ref.~\cite{B95} where the second (2PN)
correction terms in the mass-type multipole moments and the first (1PN)
ones in the current-type multipole moments were obtained.  An equivalent
expression of the 1PN current-type moments had been derived earlier
\cite{DI91a}, however in a different form which is not amenable to
generalization to higher order (but turned out to be useful in applications
\cite{BDI95}).

The main result of the present paper is the expression of the 2.5PN-accurate
energy loss by gravitational radiation from a general (semi relativistic)
source, and from an inspiralling compact binary. In the latter case
of application, the 2.5PN contribution in the energy loss is found to be
entirely due to tails in the wave zone (this is like the 1.5PN contribution),
and to reduce in the test-body limit to the known result of perturbation
theory \cite{TNaka94,Sasa94,TSasa94}.

 With the energy loss one can derive the laws of variation of the
inspiralling binary's orbital frequency and phase using an energy balance
equation. However note that this is a weak point of the analysis because the
latter energy balance equation has been proved to hold only at the Newtonian
order (see \cite{CE70,Ker80,PaL81,BRu81,A87} for general systems, and \cite{DD}  
for binary systems), and more recently at the 1PN order \cite{IW,B'}. It is
also known to hold for the specific effects of tails at 1.5PN order
\cite{BD88,BD92}. To prove this equation at the 2.5PN order as we would need
below, one should in principle obtain the equations of motion of the binary
up to the very high 5PN order (or order $c^{-10}$) beyond the Newtonian
acceleration. Indeed the radiation reaction forces which are responsible
for the decrease of the binding energy of the binary are themselves
dominantly of order 2.5PN beyond the Newtonian term. The only method which
is available presently in order to deal with this problem is to {\it assume}
that the 2.5PN-accurate radiation reaction forces are such that there is exact
agreement between the loss of 2.5PN-accurate binding energy of the binary
(as computed from the Damour-Deruelle equations of motion \cite{DD}) and the
2.5PN-accurate energy flux we shall compute below. This assumption is verified
at the 1.5PN order and sounds reasonable, but will have to be
justified in future work.

As an indication of the quantitative importance of the 2.5PN approximation
in the orbital phase of the binary, we compute the contribution of the 2.5PN
term to the number of gravitational-wave cycles between the entry and exit
frequencies of some detector. Essentially we find that the 2.5PN approximation
is of the same order of magnitude as the 2PN term (computed in \cite{BDIWWi95}),
but opposite in sign. In the case of two neutron stars of mass $1.4 M_{\odot}$
and of the frequency bandwidth [10 Hz, 1000 Hz], the 2PN term contributed to
$+9$ units to the total number of cycles \cite{BDIWWi95}. We find here that the
2.5PN term contributes to $-11$ cycles in the same conditions. This shows the
importance of the 2.5PN approximation for constructing accurate theoretical
templates. [However, we find that the contribution of the finite mass effects
in the 2.5PN term (which cannot be obtained in perturbation theory) is
numerically small.]

In the present paper we shall make a thorough investigation (see Sections~II,
III and IV below) of all the relativistic corrections in the multipole
moments of the system which contribute to the 2.5PN-accurate energy loss.
However when we are interested only in the application to inspiralling compact
binaries, this investigation can be seen a posteriori to be unnecessary.
Indeed the orbit of an inspiralling binary is {\it circular}, and we shall
prove in this case that the 2.5PN relativistic corrections in the multipole
moments give in fact no contribution in the energy loss. As we said above,
the only contribution is that of the tails present at this order (see Section
VI). A simple argument (concerning the result at 2.5PN order of a contracted
product of tensors made of the relative separation and velocity of the bodies)
could be used beforehand to see that this is true. But because inspiralling
compact binaries may not constitute the only sources for which the 2.5PN
approximation in the energy loss is needed, or simply because one may need in
the future to consider the case of a binary moving on an eccentric orbit, we
have chosen in this paper to compute systematically all the terms which enter
the 2.5PN-accurate energy loss for general systems. This permits us to show
explicitly that all the terms but the tail terms give zero in the energy loss
for (circular) inspiralling binaries. The simple argument mentioned above may
be used in future work to simplify the investigation of higher
post-Newtonian orders.

The plan of this paper is as follows. In the following section II and in the
next one III we follow step by step the derivation done in Ref.~\cite{B95}
of the near-zone gravitational field and the corresponding matching
equation, and show how this derivation can be extended to the 2.5PN order.
In section IV we obtain the explicit 2.5PN corrections arising in the
mass-type multipole moments.  Section~V deals with the derivation of the
energy loss formula valid for general systems (however some coefficients
are left un-specified in the formula). Finally these results are applied
in Section~VI to inspiralling compact binaries. Appendix A
derives a useful integration formula and Appendix B presents a relevant
summary of the 2.5PN equations of motion.

Throughout this paper we refer to Ref.~\cite{B95} as paper~I and to
Ref.~\cite{BDI95} as paper~II.

\section{The gravitational field in the near zone}
\label{sec:2}

Following the plan of paper~I we first investigate the gravitational field
generated by a slowly moving isolated source in its {\it near zone}, which
is defined in the usual way as being a zone whose size is of small extent
with respect to a typical wavelength of the emitted radiation. Two
distinctive methods are used. The first one is a direct post-Newtonian
iteration (speed of light $c\to +\infty$) of the field equations inside
the source, and is valid all-over the near zone. The second method consists
of re-expanding when $c\to +\infty$ a solution of the vacuum field equations
obtained by means of the multipolar and post-Minkowskian iteration scheme
of Ref.~\cite{BD86}, and is valid only in the exterior part of the near zone.

We denote the small post-Newtonian parameter by $\varepsilon \sim
v/c$, where $v$ is a typical velocity in the source (e.g., the relative
orbital velocity of the two bodies in the case of a binary system).  A
remainder term of order $O(\varepsilon^n )$ is abbreviated by $O(n)$.  In a
vector $A^\mu$ or a tensor $B^{\mu\nu}$, the remainder term  is denoted by
$O(n,p)$ or $O(n,p,q)$, by which we mean a term of order $O(n)$ in $A^0$ or
$B^{00}$, of order $O(p)$ in $A^i$ or $B^{0i}$ or $B^{i0}$, and of order
$O(q)$ in $B^{ij}$.  [Greek indices range from 0 to 3, and Latin indices
range from 1 to 3.] Most of the notations used here are like in paper~I.

\subsection{The inner gravitational field}
\label{sec:2.1}

$T^{\alpha\beta}$ are the contravariant components of the stress-energy
tensor (with dimension of an energy density) of the material source in some
inner coordinate system $({\bf x},t)$. The densities of mass $\sigma$, of
current $\sigma_i$ and of stress $\sigma_{ij}$ in the source are defined by
\begin{subequations}
\label{eq:2.1}
\begin{eqnarray}
 \sigma &=& {T^{00} + T^{ii} \over c^2}\ , \label{eq:2.1a}\\
 \sigma_i &=& {T^{0i} \over c}\ , \label{eq:2.1b}\\
 \sigma_{ij} &=& T^{ij} \ , \label{eq:2.1c}
\end{eqnarray}
\end{subequations}
where $T^{ii}$ denotes the spatial trace $\Sigma \delta_{ij} T^{ij}$. These
definitions are such that $\sigma$, $\sigma_i$, $\sigma_{ij}$ have a
finite non-zero limit as $c\to +\infty$. From these matter densities one
defines the retarded potentials
\begin{subequations}
\label{eq:2.2}
\begin{eqnarray}
 V &=& -4\pi\,G\,\Box^{-1}_R \sigma \ , \label{eq:2.2a} \\
 V_i &=& -4\pi\,G\,\Box^{-1}_R \sigma_i \ ,\label{eq:2.2b} \\
 W_{ij} &=& -4\pi\,G\,\Box^{-1}_R \left[ \sigma_{ij} + {1\over 4\pi G}
 \left( \partial_i V \partial_j V - {1\over 2} \delta_{ij} \partial_k V
  \partial_k V \right) \right] \ ,  \label{eq:2.2c}
\end{eqnarray}
\end{subequations}
where $G$ is Newton's constant, and where $\Box^{-1}_R$ denotes the
retarded integral operator
\begin{equation}
 (\Box^{-1}_R f)({\bf x},t) = -{1\over 4\pi} \int\!\!\!\int\!\!\!\int
 {d^3{\bf x}'\over
 |{\bf x}-{\bf x}'|} f ({\bf x}', t-{1\over c} |{\bf x}-{\bf x}'|)\ .
\label{eq:2.3}
\end{equation}
Contrarily to the sources of $V$ and $V_i$
which are of compact support:  $\Box V =-4\pi G\sigma$ and $\Box V_i
=-4\pi G\sigma_i$ where $\Box$ is the d'Alembertian operator, the source
of the potential $W_{ij}$ is not of compact support:
$\Box W_{ ij} = -4\pi G\sigma_{ij} -
\partial_i V \partial_j V + {1\over 2} \delta_{ij} \partial_k V
\partial_k V$.  Indeed we have included in $W_{ij}$ the
stress density of the (Newtonian) gravitational field itself since it
is of the same order as $\sigma_{ij}$ when $c\to +\infty$.  [Note that
paper~II used the notation $W_{ij}$ for a closely related but
different potential; here we do not follow paper~II but stick
to the notation of paper~I.] To Newtonian order the densities and potentials
so defined satisfy the equations of continuity and of motion
\begin{subequations}
\label{eq:2.4}
\begin{eqnarray}
 \partial_t \sigma + \partial_i \sigma_i &=& O(2)\ , \label{eq:2.4a}\\
 \partial_t \sigma_i+ \partial_j \sigma_{ij} &=& \sigma \partial_i V
    + O(2)\ . \label{eq:2.4b}
\end{eqnarray}
\end{subequations}
From these dynamical equations one deduces the differential identities
\begin{subequations}
\label{eq:2.5}
\begin{eqnarray}
 \partial_t V + \partial_i V_i &=& O(2)\ , \label{eq:2.5a}\\
 \partial_t V_i + \partial_j W_{ij} &=& O(2)\ . \label{eq:2.5b}
\end{eqnarray}
\end{subequations}
With the introduction in paper~I of the retarded potentials $V$, $V_i$,
$W_{ij}$, a simple expression of the gravitational field $h^{\alpha\beta}$
inside the source which is valid to some intermediate accuracy
$O(6,5,6)$ was written, namely
\begin{subequations}
\label{eq:2.6}
\begin{eqnarray}
 h^{00} &=& -{4\over c^2} V +{4\over c^4} (W_{ii} - 2V^2) +O(6)\ ,
    \label{eq:2.6a}\\
 h^{0i} &=& -{4\over c^3} V_i +O(5)\ ,\label{eq:2.6b}\\
 h^{ij} &=& -{4\over c^4} W_{ij} +O(6)\ .\label{eq:2.6c}
\end{eqnarray}
\end{subequations}
The field variable is $h^{\alpha\beta} \equiv \sqrt{-g}
g^{\alpha\beta} - \eta^{\alpha\beta}$, where $g$ and $g^{\alpha\beta}$
are the determinant and inverse of the usual covariant metric
$g_{\alpha\beta}$, and where $\eta^{\alpha\beta}$ is the Minkowski
metric (signature $-+++$). Note the important fact that there are no
{\it explicit} terms in Eqs.~(\ref{eq:2.6}) involving powers of
$c^{-1}$ which are ``odd'' in the post-Newtonian sense (e.g., a
term of order $\sim c^{-5}$ in $h^{00}$ or $h^ {ij}$). This is because
we have kept the potentials $V$, $V_i$ and $W_{ij}$ in retarded form,
without expanding when $c\to +\infty$ the retardation they contain.
The ``odd" terms in Eqs.~(\ref{eq:2.6}) could be easily computed using
the post-Newtonian expansions of the retarded potentials as given by
Eqs.~(\ref{eq:4.4}) below.

Paper~I iterated the inner field (\ref{eq:2.6}) from this
intermediate post-Newtonian order to the next order with the result that
the field to the higher precision $O(8,7,8)$ could be written as
\begin{equation}
 h^{\alpha\beta} = \Box^{-1}_{R} \left[ {16\pi G\over c^4}\,
 \overline\lambda (V,W) T^{\alpha\beta} + \overline\Lambda^{\alpha\beta}
  (V,W)\right]  + O(8,7,8)\  \label{eq:2.7}
\end{equation}
where $\overline\lambda$ and $\overline\Lambda^{\alpha\beta}$ denote
some explicit combinations of the inner gravitational potentials $V$,
$V_i$, $W_{ij}$ and their derivatives.  First we have
\begin{equation}
 \overline\lambda (V, W) = 1 + {4\over c^2} V - {8\over c^4} (W_{ii}-V^2)\
 \label{eq:2.8}
\end{equation}
which represents in fact the post-Newtonian expansion of (minus) the
determinant of the metric, with terms $O(6)$ suppressed. Second
the components of $\overline\Lambda^{\alpha\beta}$ read
\begin{subequations}
\label{eq:2.9}
\begin{eqnarray}
 \overline\Lambda^{00}(V,W) &=& - {14\over c^4} \partial_k V
 \partial_k V + {16\over c^6} \biggl\{ - V \partial^2_t V - 2V_k \partial_t
 \partial_k V \nonumber \\
 && \qquad - W_{km} \partial_{km}^2 V + {5\over 8} (\partial_t V)^2
 + {1\over 2} \partial_k V_m (\partial_k V_m +3\partial_m V_k) \nonumber\\
 && \qquad + \partial_k V\partial_t V_k + 2\partial_k V\partial_k W_{mm}
  - {7\over 2} V\partial_k V\partial_k V \biggr\} \ , \label{eq:2.9a} \\
 \overline\Lambda^{0i}(V,W) &=&  {16\over c^5} \left\{ \partial_k V
 (\partial_i V_k - \partial_k V_i) + {3\over 4} \partial_t V \partial_i V
 \right\} \ , \label{eq:2.9b}\\
 \overline\Lambda^{ij}(V,W) &=& {4\over c^4}\left\{ \partial_i V
 \partial_j V -{1\over 2} \delta_{ij} \partial_k V \partial_k V \right\}
  + {16\over c^6} \biggl\{ 2 \partial_{(i} V\partial_t V_{j)}
  - \partial_i V_k \partial_j V_k  \nonumber \\
 && \qquad - \partial_k V_i \partial_k V_j
  + 2 \partial_{(i} V_k \partial_k V_{j)} - {3\over 8} \delta_{ij}
    (\partial_t V)^2 - \delta_{ij} \partial_k V \partial_tV_k \nonumber \\
 && \qquad + {1\over 2} \delta_{ij}\partial_k V_m (\partial_k V_m
   - \partial_m V_k) \biggr\} \ , \label{eq:2.9c}
\end{eqnarray}
\end{subequations}
and represent the expansion of the effective nonlinear gravitational
source of Einstein's equations in harmonic coordinates, with $O(8,7,8)$
terms suppressed.  The overbar on
$\overline\lambda$ and $\overline\Lambda^{\alpha\beta}$ reminds us that
these quantities are only determined up to a certain post-Newtonian
order.  The (approximate) harmonic coordinate condition is
\begin{equation}
 \partial_\beta h^{\alpha\beta} = O(7,8)\ . \label{eq:2.10}
\end{equation}
As we shall see the post-Newtonian accuracy of the inner
field (\ref{eq:2.7})-(\ref{eq:2.10}) is sufficient for our purpose.

\subsection{The external gravitational field}
\label{sec:2.2}

In the exterior we use a solution of the vacuum field equations which
has in principle sufficient generality for dealing with an arbitrary
source of gravitational radiation.  This solution is given as a
(nonlinear) functional of two infinite sets of time-varying multipole
moments, $M_L(t)$ and $S_L(t)$.  The index $L$ carried by these moments
represents a multi-index formed with $\ell$ spatial indices:
$L=i_1i_2\dots i_\ell$.  The ``order of multipolarity'' $\ell$ goes from
zero to infinity for the ``mass-type'' moments $M_L(t)$ and from one to
infinity for the ``current-type'' moments $S_L(t)$.  The moments $M_L$
and $S_L$ are symmetric and tracefree (STF) in their $\ell$ indices.
The mass monopole $M$ is simply the total mass of the source or ADM
mass, the mass dipole $M_i$ is the position of the center of mass (in
units of total mass), and the current dipole $S_i$ is the total angular
momentum.  $M$, $M_i$ and $S_i$ are constant. Furthermore
we shall choose $M_i=0$ by translating the spatial origin of the
coordinates to the center of mass.

Some external potentials playing a role analogous to the
inner potentials but differing from them both in structural form and
numerical values are introduced.  First the potentials $V^{\rm ext}$,
$V^{\rm ext}_i$ and $V^{\rm ext}_{ij}$ are given by their
multipolar series parametrized by the multipole moments $M_L(t)$ and
$S_L(t)$~:
\begin{subequations}
\label{eq:2.11}
\begin{eqnarray}
 V^{\rm ext} &=& G \sum^\infty_{\ell =0} {(-)^\ell\over \ell !}
\partial_L \left[ {1\over r} \,M_L \left( t-{r\over c}\right)\right]\ ,
 \label{eq:2.11a} \\
 V_i^{\rm ext} &=& -G \sum^\infty_{\ell =1} {(-)^\ell\over \ell !}
  \partial_{L-1} \left[ {1\over r} \,M^{(1)}_{iL-1}
  \left( t-{r\over c}\right)\right]\nonumber \\
   && -G \sum^\infty_{\ell =1} {(-)^\ell\over \ell !}{\ell\over \ell +1}
 \varepsilon_{iab} \partial_{aL-1} \left[ {1\over r} \,S_{bL-1}
  \left( t-{r\over c}\right)\right]\ , \label{eq:2.11b}\\
 V_{ij}^{\rm ext} &=& G \sum^\infty_{\ell =2} {(-)^\ell\over \ell !}
  \partial_{L-2} \left[ {1\over r} \,M^{(2)}_{ijL-2}
  \left( t-{r\over c}\right)\right]\nonumber \\
   && +G \sum^\infty_{\ell =2} {(-)^\ell\over \ell !}{2\ell\over \ell +1}
\partial_{aL-2} \left[ {1\over r} \varepsilon_{ab(i}  \,S^{(1)}_{j)bL-2}
  \left( t-{r\over c}\right)\right]\ . \label{eq:2.11c}
\end{eqnarray}
\end{subequations}
The notation $\partial_L$ is shorthand for a product of partial derivatives,
$\partial_L =\partial_{i_1} \partial_{i_2}\dots \partial_{i_\ell}$ where
$\partial_i =\partial /\partial x^i$. In a similar way
$\partial_{L-1} =\partial_{i_1}\dots \partial_{i_{\ell-1}}$, $\partial_{aL-1}
=\partial_a \partial_{L-1}$, and so on.  The superscript $(n)$ indicates
$n$ time derivatives, and the indices in brackets are symmetrized.  The
potentials (\ref{eq:2.11}) satisfy the source-free d'Alembertian
equation and the (exact) differential identities
\begin{subequations}
\label{eq:2.12}
\begin{eqnarray}
\partial_t V^{\rm ext}+\partial_i V_i^{\rm ext} &=& 0\ ,\label{eq:2.12a}\\
\partial_t V_i^{\rm ext}+\partial_j V_{ij}^{\rm ext} &=& 0\ .\label{eq:2.12b}
\end{eqnarray}
\end{subequations}
Furthermore $V^{\rm ext}_{ij}$ is tracefree:  $V^{\rm ext}_{ii}=0$.
Having defined these potentials one introduces a more
complicated potential $W^{\rm ext}_{ij}$ by the formula
\begin{equation}
 W^{\rm ext}_{ij} = V^{\rm ext}_{ij} + \hbox{FP}_{B=0}\ \Box^{-1}_R
  \left[ r^B \left( -\partial_i V^{\rm ext} \partial_j V^{\rm ext}
 + {1\over 2} \delta_{ij} \partial_k V^{\rm ext} \partial_k
   V^{\rm ext} \right) \right]\ . \label{eq:2.13}
\end{equation}
The second term appearing here involves the retarded integral (\ref{eq:2.3})
but regularized by means of the analytic continuation process defined in
Ref.~\cite{BD86}. One multiplies the source by $r^B$ where $r=|{\bf x}|$
and $B$ is a complex number. Then applying the operator $\Box^{-1}_R$
defines a function of $B$ which can be analytically continued all-over the
complex plane deprived of the integers, but admitting there a Laurent
expansion with only some simple poles \cite{BD86}. The looked-for solution
is equal to the finite part at $B=0$ (in short FP$_{B=0}$) or constant
term $\sim B^0$ in the Laurent expansion when $B \to 0$.  This
regularization process is made indispensable by the fact that we are
looking for solutions of the wave equation in the form of multipole
expansions, which are valid only in the exterior of the source and are
singular when considered formally inside the source.

With the external potentials $V^{\rm ext}$, $V^{\rm ext}_i$ and
$W_{ij}^{\rm ext}$ one has a result similar to Eq.~(\ref{eq:2.6}), namely
the expression of the external field $h^{\mu\nu}_{\rm can}$ (where
the notation ``can'' means that we are considering specifically the
``canonical'' construction of the external field as defined in Sect.~4.3
of \cite{BD86}) up to the intermediate accuracy $O(6,5,6)$~:
\begin{subequations}
\label{eq:2.14}
\begin{eqnarray}
 h^{00}_{\rm can}&=& - {4\over c^2}\ V^{\rm ext} + {4\over c^4}\
 (W_{ii}^{\rm ext} - 2 (V^{\rm ext})^2) + O(6)\ , \label{eq:2.14a} \\
 h^{0i}_{\rm can}&=& -{4\over c^3}\ V^{\rm ext}_i+ O(5)\ ,\label{eq:2.14b}\\
 h^{ij}_{\rm can}&=& -{4\over c^4}\ W^{\rm ext}_{ij}+O(6)\ .\label{eq:2.14c}
\end{eqnarray}
\end{subequations}
Because this expression has the same form as Eq.~(\ref{eq:2.6}), the
nonlinearities in the exterior field will have in turn the same form
as Eq.~(\ref{eq:2.9}).  Up to order $O(8,7,8)$ we obtain the
``canonical" field
\begin{equation}
 h^{\mu\nu}_{\rm can} = Gh^{\mu\nu}_{\rm can(1)}+G^2 q^{\mu\nu}_{\rm can(2)}
  + {\rm FP}_{B=0} \ \Box^{-1}_R \left[ r^B \overline\Lambda^{\mu\nu}
  (V^{\rm ext}, W^{\rm ext})\right] + O(8,7,8)\ ,
\label{eq:2.15}
\end{equation}
satisfying the (exact) harmonic gauge condition
\begin{equation}
  \partial_\nu h^{\mu\nu}_{\rm can} = 0\ . \label{eq:2.16}
\end{equation}
In the last term of (\ref{eq:2.15}) the symbol FP$_{B=0}$ and the
regularization
factor $r^B$ have the same meaning as in (\ref{eq:2.13}). The effective
nonlinear source $\overline\Lambda^{\mu\nu}$ is given by Eqs.~(\ref{eq:2.9})
but expressed with the external potentials $V^{\rm ext}$, $V^{\rm ext}_i
$ and $W^{\rm ext}_{ij}$ instead of the inner potentials $V$, $V_i$ and
$W_{ij}$ appearing in (\ref{eq:2.7}).  The first term
$Gh^{\mu\nu}_{\rm can(1)}$ in Eq.~(\ref{eq:2.15}) is linear in the
external potentials,
\begin{subequations}
\label{eq:2.17}
\begin{eqnarray}
 G\,h^{00}_{\rm can(1)} &=&- {4\over c^2}\,V^{\rm ext}\ ,\label{eq:2.17a}\\
 G\,h^{0i}_{\rm can(1)} &=&- {4\over c^3}\,V_i^{\rm ext}\ ,\label{eq:2.17b}\\
 G\,h^{ij}_{\rm can(1)} &=&- {4\over c^4}\,V_{ij}^{\rm ext}\ .
     \label{eq:2.17c}
\end{eqnarray}
\end{subequations}
This term is the solution of the linearized (vacuum) equations on which
is based the post-Minkowskian algorithm for the construction of the
canonical metric in Ref.~\cite{BD86}. The powers of $c^{-1}$ in Eqs.
(\ref{eq:2.17}) are such that $h^{\mu\nu}_{\rm can(1)}= O(2,3,4)$. Finally
the term $G^2q^{\mu\nu}_{\rm can(2)}$ in (\ref{eq:2.15}) is a particular
solution of the source-free wave equation which has to be added
in order that the harmonic gauge condition (\ref{eq:2.16}) be satisfied.
It was proved in the Appendix~A of paper~I that this term is of order
\begin{equation}
 G^2 q^{\mu\nu}_{\rm can(2)} = O(7,7,7)\ , \label{eq:2.18}
\end{equation}
and thus can be safely neglected if we are interested in the mass multipole
moments to 2PN order only.  In the present work investigating the 2.5PN
order the term (\ref{eq:2.18}) cannot {\it a priori} be neglected. However
we shall see that the 2.5PN order is needed only in the sum of the 00 component
and the spatial trace of this term, i.e.  $G^2 (q^{00}_{\rm can(2)} +
q^{ii}_{\rm can(2)}$). Relying on our previous papers we know that this sum
is made of some retarded waves of the type $\partial_L (r^{-1} X (t-r/c))$
with scalar or dipolar multipolarity only $(\ell=0$ or 1). Furthermore we
know that the dependence on $c^{-1}$ of such a wave is $O(5+\underline\ell_1
+ \underline\ell_2 -\ell)$ where $\underline\ell_1$ and $\underline\ell_2$ are
the multipolarities of the two interacting moments composing the wave (we use
the same notation as e.g. the Appendix~A of paper~I; see also after
Eq.~(\ref{eq:5.5}) below).  With $\ell=0$ or 1 a term of order $O(7)$
necessarily has $\underline\ell_1 +\underline\ell_2 = 2$ or 3. The term in
question is made of the product of the mass $M$ with the quadrupoles $M_{ij}$
and $S_{ij}$, or of the product of the mass dipole $M_i$ with the quadrupole
$M_{ij}$. The first possibility is excluded because $M_{ij}$ and $S_{ij}$ are
STF (thus no scalar or dipole wave with no free index can be formed), and the
second possibility does not exist in a mass centred-frame where $M_i=0$.
So we have proved that in a mass-centred frame we have
\begin{equation}
 G^2 \left(q^{00}_{\rm can(2)} + q^{ii}_{\rm can(2)}\right)
       = O(8)\ ,  \label{eq:2.19}
\end{equation}
which is all what will be needed in the following.

\section{The matching between the inner and outer fields}
\label{sec:3}

\subsection{Relations between inner and outer potentials}
\label{sec:3.1}

Our matching requirement is that there exists a change of coordinates
valid in the external near zone and transforming the inner
gravitational field $h^{\alpha\beta} (x)$ given by Eq.~(\ref{eq:2.7})
into the outer field $h^{\alpha\beta}_{\rm can} (x_{\rm can})$ given by
Eq.~(\ref{eq:2.15}).  Let this change of coordinates be
\begin{equation}
  x^\mu_{\rm can} (x) = x^\mu + \varphi^\mu (x) \label{eq:3.1}
\end{equation}
where $x^\mu$ are the inner coordinates used in section II.A and
$x^\mu_{\rm can}$ the outer coordinates used in section II.B. The vector
$\varphi^\mu$ has been shown in paper I to be of order
\begin{equation}
  \varphi^\mu = O(3,4)\ . \label{eq:3.2}
\end{equation}
Because the two coordinate systems
$x^\mu$ and $x^\mu_{\rm can}$ are harmonic [at least approximately~; see
Eqs.~(\ref{eq:2.10}) and (\ref{eq:2.16})], and because $h^{\alpha\beta}
= O(2,3,4)$ in addition to Eq.~(\ref{eq:3.2}), we have
\begin{equation}
 \Box \varphi^\mu = O(7,8)\ . \label{eq:3.3}
\end{equation}
The matching equations consistent with the order $O(8,7,8)$
read from paper~I as
\begin{subequations}
\label{eq:3.4}
\begin{eqnarray}
 h^{00}_{\rm can} (x) &=& h^{00} (x) + \partial\varphi^{00}
  + 2h^{0\mu}\partial_\mu \varphi^0 - \partial_\mu (h^{00}\varphi^\mu)
  + \partial_i\,\varphi^0\,\partial_i\,\varphi^0 + O(8)\ ,\label{eq:3.4a}\\
 h^{0i}_{\rm can} (x) &=& h^{0i} (x) + \partial\varphi^{0i}
  + O(7)\ , \label{eq:3.4b}\\
 h^{ij}_{\rm can} (x) &=& h^{ij} (x) + \partial\varphi^{ij}
  + O(8)\ , \label{eq:3.4c}
\end{eqnarray}
\end{subequations}
where both the inner and outer fields are expressed in the same (inner)
coordinate system $x^\mu$.  We denote by $\partial\varphi^{\mu\nu}$ the
linear part of the coordinate transformation,
\begin{equation}
\partial\varphi^{\mu\nu} =\partial^\mu\varphi^\nu + \partial^\nu\varphi^\mu
   - \eta^{\mu\nu} \partial_\lambda \varphi^\lambda\ . \label{eq:3.5}
\end{equation}
The nonlinear part of the transformation enters to this order only the 00
component (\ref{eq:3.4a}).  The matching equations were used in paper~I
first to obtain the relations between the external potentials and the
{\it multipole} expansions of the corresponding inner potentials.  The
insertion of the intermediate expressions of $ h^{\mu\nu}$ and
$h^{\mu\nu}_{\rm can}$ valid up to $O(6,5,6)$ and given by (\ref{eq:2.6}) and
(\ref{eq:2.14}) yields
\begin{subequations}
\label{eq:3.6}
\begin{eqnarray}
V^{\rm ext} &=& {\cal M}(V)+c\,\partial_t\varphi^0 + O(4)\ ,\label{eq:3.6a}\\
V^{\rm ext}_i &=& {\cal M}(V_i) -{c^3\over 4}\,\partial_i\,\varphi^0 + O(2)
    \ , \label{eq:3.6b}\\
 W^{\rm ext}_{ij} &=& {\cal M}(W_{ij}) -{c^4\over 4}
  [\partial_i \varphi^j + \partial_j \varphi^i - \delta_{ij} (\partial_0
  \varphi^0 +\partial_k \varphi^k) ] + O(2)\ .     \label{eq:3.6c}
\end{eqnarray}
\end{subequations}
The script letter ${\cal M}$ refers to the multipole expansion. Note that
the first equation is valid to post-Newtonian order [see the remainder $O(4)$]
while the two others are valid to Newtonian order only [remainder
$O(2)$]. The multipole expansions of $V$, $V_i$ and $W_{ij}$ are given by
\begin{subequations}
\label{eq:3.7}
\begin{eqnarray}
 {\cal M}(V) &=& G \sum^{+\infty}_{\ell=0} {(-)^\ell\over \ell !}
 \partial_L \left[ {1\over r} {\cal V}^L \left( t-{r\over c}\right)\right]
  \ , \label{eq:3.7a} \\
 {\cal M}(V_i) &=& G \sum^{+\infty}_{\ell=0} {(-)^\ell\over \ell !}
 \partial_L \left[ {1\over r} {\cal V}^L_i \left( t-{r\over c}\right)\right]
  \ , \label{eq:3.7b} \\
 {\cal M}(W_{ij}) &=& G \sum^{+\infty}_{\ell=0} {(-)^\ell\over \ell !}
 \partial_L \left[{1\over r}{\cal W}^L_{ij}\left( t-{r\over c}\right)\right]
  \nonumber\\
 &&+ {\rm FP}_{B=0} \Box^{-1}_R \left[ r^B \left( -\partial_i {\cal M}(V)
  \partial_j {\cal M}(V) + {1\over 2} \delta_{ij} \partial_k {\cal M}(V)
  \partial_k {\cal M}(V) \right) \right] \ , \label{eq:3.7c}
\end{eqnarray}
\end{subequations}
where the (reducible) multipole moments ${\cal V}^{L}$, ${\cal V}_i^{L}$
and ${\cal W}_{ij}^{L}$ are
\begin{subequations}
\label{eq:3.8}
\begin{eqnarray}
 {\cal V}^L(t) &=& \int d^3{\bf x}\,\hat x_L \int^1_{-1} dz\, \delta_\ell
 (z)\sigma ({\bf x},t+z |{\bf x}|/c)\ ,\label{eq:3.8a}\\
 {\cal V}^L_i(t) &=& \int d^3{\bf x}\,\hat x_L \int^1_{-1} dz\, \delta_\ell
 (z)\sigma_i ({\bf x},t+z |{\bf x}|/c)\ ,\label{eq:3.8b} \\
 {\cal W}_{ij}^L(t) &=& {\rm FP}_{B=0} \int d^3 {\bf x}\,
  |{\bf x}|^B \hat x_L \int^1_{-1} dz\, \delta_\ell (z) \nonumber \\
 &&\times \left[ \sigma_{ij} +{1\over 4\pi G} \left(\partial_i V\partial_j V
   - {1\over 2} \delta_{ij} \partial_k V\partial_k
    V\right)\right] ({\bf x}, t+z |{\bf x}|/ c)\ . \label{eq:3.8c}
\end{eqnarray}
\end{subequations}
The notation $\hat x_L$ is for the tracefree projection of the product
of $\ell$ spatial vectors $x_L\equiv x_{i_1}\cdots x_{i_\ell}$.
The function $\delta_\ell (z)$ is given by
\begin{equation}
\delta_\ell (z) ={(2\ell +1)!!\over 2^{\ell+1}\ell !} (1-z^2)^\ell\quad ;
 \qquad \quad \int^1_{-1} dz\, \delta_\ell (z) =1\ . \label{eq:3.9}
\end{equation}
Some explanations on the expressions (\ref{eq:3.7})-(\ref{eq:3.9}) are in
order. First notice that the expressions of the multipole expansions of the
potentials $V$ and $V_i$ whose sources have a compact support are quite
standard. They can be found in this form in the Appendix~B of Ref.~\cite{BD89}
(but were derived earlier in an alternative form \cite{CMM77}). Notably the
presence of the function $\delta_\ell (z)$ is due to the time
delays of the propagation of the waves with finite velocity $c$ inside the
source. By contrast the expression of the multipole expansion of the
potential $W_{ij}$ whose source extends up everywhere in space is more
complicated. The second term in Eq.~(\ref{eq:3.7c}) ensures that
${\cal M}(W_{ij})$ satisfies the correct equation [deduced from
(\ref{eq:2.2c})] outside the source, namely $\Box {\cal M}(W_{ij})
=-\partial_i {\cal M}(V) \partial_j{\cal M}(V) + {1\over 2} \delta_{ij}
\partial_k {\cal M}(V) \partial_k {\cal M}(V)$ where the right side
agrees numerically with $-\partial_i V\partial_j V + {1\over 2} \delta_{ij}
\partial_k V \partial_k V$ outside the source. This second term involves
the multipole expansion ${\cal M}(V)$ of the inner potential and not $V$
itself. This is in conformity with the use of the regularized operator
FP $\Box^{-1}_R$ which is defined only when acting on multipole expansions
(like in Eq.~(\ref{eq:2.13})).  On the other hand the integrand of
(\ref{eq:3.8c}) involves the non-compact supported source of $W_{ij}$
where appears the potential $V$ itself, i.e.  not in multipole expanded
form.  Thus the integrand of (\ref{eq:3.8c}) is valid everywhere inside
and outside the source.  Very far from the source it diverges because of
the presence of the product $\hat x_L$ of $\ell$ spatial vectors, behaving
like $|{\bf x}|^\ell$ at spatial infinity.  The well-definiteness of the
integral results from the presence of the regularization factor $|{\bf x}|^B$
and the finite part symbol. We notice that no {\it ad hoc} prescription is
necessary in order to obtain the multipole moments in a well-defined form
even in the case of a non-compact supported source. This is proved in paper~I.

\subsection{The matching equation} 
\label{sec:3.2}

The relations (\ref{eq:3.6}) linking $V^{\rm ext}$, $V^{\rm ext}_i$,
$W^{\rm ext}_{ij}$ to the multipole expansions ${\cal M}(V)$, ${\cal M}(V_i)$,
${\cal M}(W_{ij})$ serve us with re-expressing the non-linearities in
the external metric in terms of the potentials belonging to the inner
metric.  The result of paper~I is
\begin{equation}
\overline{\Lambda}^{\mu\nu} (V^{\rm ext}, W^{\rm ext})
 =\overline{\Lambda}^{\mu\nu} ({\cal M}(V),{\cal M}(W)) +
 \Box \Omega^{\mu\nu} + O(8,7,8)\ , \label{eq:3.10}
\end{equation}
where the components of the tensor $\Omega^{\mu\nu}$ are given by
\begin{subequations}
\label{eq:3.11}
\begin{eqnarray}
 \Omega^{00} &=& - {8\over c^3}\, \left[ {\cal M}(V) \partial_t \varphi^0
  + {\cal M}(V_i) \partial_i\varphi^0 - {c\over 2} \partial_\mu
  ({\cal M} (V) \varphi^\mu)\right] + \partial_i \varphi^0
   \partial_i\varphi^0\ , \label{eq:3.11a}\\
 \Omega^{0i} &=& \Omega^{ij} = 0\ . \label{eq:3.11b}
\end{eqnarray}
\end{subequations}
The only non-zero component of this tensor is the 00 component which
is of order
\begin{equation}
 \Omega^{00} = O(6)\ .  \label{eq:3.12}
\end{equation}
This is because $\varphi^0 =O(3)$, $\varphi^i=O(4)$.  Both sides of
Eq.~(\ref{eq:3.10}) are multiplied by $r^B$, and one applies afterwards
the retarded integral $\Box^{-1}_R$ and take the finite part at $B = 0$.
We find
\begin{equation}
 {\rm FP}_{B=0} \Box^{-1}_R [r^B \overline\Lambda^{\mu\nu} (V^{\rm ext},
 W^{\rm ext})] = {\rm FP}_{B=0} \Box^{-1}_R [r^B \overline\Lambda^{\mu\nu}
 ({\cal M}(V), {\cal M}(W))] + \Omega^{\mu\nu} + X^{\mu\nu} +
 O(8,7,8)   \label{eq:3.13}
\end{equation}
where one must be careful that an extra term $X^{\mu\nu}$ with respect to
paper~I arises whose components are given by
\begin{subequations}
\label{eq:3.14}
\begin{eqnarray}
  X^{00} &=& {\rm FP}_{B=0}\, \Box^{-1}_R [r^B \Box \Omega^{00}]
    - \Omega^{00} \ , \label{eq:3.14a}\\
 X^{0i} &=& X^{ij} = 0\ . \label{eq:3.14b}
\end{eqnarray}
\end{subequations}
The only non-zero component of this term is the 00 component which satisfies
\begin{equation}
 \Box X^{00} = 0\quad ; \qquad\quad X^{00} = O(7)\ . \label{eq:3.15}
\end{equation}
The fact that $X^{00}$ is a (retarded) solution of the source-free d'Alembert
equation is clear from its definition (\ref{eq:3.14a}) and the main
property of the operator FP$\Box^{-1}_R$ which is the inverse of $\Box$
(when acting on multipolar sources). The fact that $X^{00}$ is of order
$O(7)$, that is exactly one order in $c^{-1}$ smaller than the order of
the corresponding $\Omega^{00}$, is not immediately obvious but will be
proved below.  The quantity $X^{00}$ was rightly neglected in paper~I
but has to be considered here since it will contribute to the mass
moments at the 2.5PN order.

We are now in a position to write down a matching equation valid up to the
neglect of $O(8,7,8)$ terms. The equation is obtained by insertion into
Eqs.~(\ref{eq:3.4}) of both the inner field $h^{\mu\nu}$ given by
(\ref{eq:2.7}) and the outer field $h^{\mu\nu}_{\rm can}$ given by
(\ref{eq:2.15}). Use is made of the link derived in (\ref{eq:3.13}) between
the external and internal non-linearities. We find that $\Omega^{\mu\nu}$
cancels to the required order the nonlinear part of the coordinate
transformation so that only remains the linear part $\partial\varphi^{\mu\nu}$
given by Eq.~(\ref{eq:3.5}).  The resulting matching equation (extending
the less accurate equation (3.35) of paper I) reads as
\begin{eqnarray}
 Gh^{\mu\nu}_{\rm can(1)} + G^2 q^{\mu\nu}_{\rm can(2)}
 =&& \Box^{-1}_R \left[ {16\pi G\over c^4}\, \overline\lambda (V,W) T^{\mu\nu}
  +\overline\Lambda^{\mu\nu} (V,W) \right] \nonumber \\
 &&- {\rm FP}_{B=0}\ \Box^{-1}_R [r^B\, \overline\Lambda
  ^{\mu\nu} ({\cal M} (V), {\cal M}(W))] \nonumber \\
 &&- X^{\mu\nu} + \partial\varphi^{\mu\nu} + O(8,7,8)\ . \label{eq:3.16}
\end{eqnarray}
There are two new terms with respect to (3.35) in paper~I~:
$G^2q^{\mu\nu}_{\rm can(2)}$ in the left side and $-X^{\mu\nu}$ in the
right side which both are terms of 2.5PN order.  Now, by exactly the
same reasoning as in paper~I one can transform the difference between
the two retarded integrals in the right side into an explicit multipole
expansion parametrized by some moments $\overline{\cal
T}^{\mu\nu}_L(t)$, namely
\begin{equation}
 Gh^{\mu\nu}_{\rm can(1)}+ G^2 q^{\mu\nu}_{\rm can(2)}
  = -{4G\over c^4} \sum^{+\infty}_{\ell =0}
 {(-)^\ell\over \ell !} \partial_L \left[ {1\over r}\,
  \overline{\cal T}^{\mu\nu}_L (t-r/c)\right] - X^{\mu\nu} +
  \partial\varphi^{\mu\nu} + O(8,7,8)\ . \label{eq:3.17}
\end{equation}
These moments are given by
\begin{equation}
 \overline{\cal T}^{\mu\nu}_L(t) = {\rm FP}_{B=0} \int
  d^3 {\bf x}\,|{\bf x}|^B \hat x_L \int^1_{-1} dz\, \delta_\ell (z)
 \overline\tau^{\mu\nu} ({\bf x}, t+z |{\bf x}|/c)\   \label{eq:3.18}
\end{equation}
where $\overline\tau^{\mu\nu}$ denotes the total stress-energy tensor of
the material and gravitational fields (valid up to the
considered precision)
\begin{equation}
 \overline\tau^{\mu\nu}(V,W) = \overline\lambda (V,W) T^{\mu\nu}
  + {c^4\over 16\pi G}\overline\Lambda^{\mu\nu} (V,W) \ ,
    \label{eq:3.19}
\end{equation}
which is conserved in the sense
\begin{equation}
 \partial_\nu \overline\tau^{\mu\nu} = O(3,4)\ .  \label{eq:3.20}
\end{equation}
It is important to note that the effective stress-energy tensor
$\overline\tau^{\mu\nu}$ is a functional of the potentials $V$, $V_i$
and $W_{ij}$ valid everywhere inside and outside the source.  There is no
contribution of the multipole expansions of the potentials
in the final result (see paper~I).

The left side of the matching equation (\ref{eq:3.17}) is a functional of
the original multipole moments $M_L$, $S_L$ parametrizing the exterior
metric.  On the other hand the right side is a functional of the actual
densities of mass, current and stress of the material fields in the
source. To find the explicit expressions of $M_L$ and $S_L$ in terms of
these source densities we decompose the right side into irreducible
multipole moments. Inspection of the reasoning done in section IV.A of
paper~I shows that this reasoning is still valid in the present more
accurate case. As a result we find
\begin{equation}
 Gh^{\mu\nu}_{\rm can(1)} [M_L,S_L] + G^2 q^{\mu\nu}_{\rm can(2)} =
 Gh^{\mu\nu}_{\rm can(1)} [I_L,J_L] - X^{\mu\nu} + \partial\xi^{\mu\nu}
 + O(8,7,8)\ , \label{eq:3.21}
\end{equation}
where the last term is a linear coordinate transformation
associated with the vector $\xi^\mu =\varphi^\mu+\omega^\mu$ where
$\omega^\mu$ is the same as in Eq.~(4.4) of paper~I.  The linearized
metric $Gh^{\mu\nu}_{\rm can(1)}$ in the right side takes the same
expression as in the left side but is parametrized instead of $M_L$ and
$S_L$ by the (STF) {\it source} multipole moments $I_L$ and $J_L$ given by
\begin{subequations}
\label{eq:3.22}
\begin{eqnarray}
 I_L(t) &=& {\rm FP}_{B=0}\int d^3 {\bf x}|{\bf x}|^B
 \int^1_{-1} dz \biggl[ \delta_\ell (z) \hat x_L \overline\Sigma -
 {4(2\ell +1)\over c^2(\ell +1)(2\ell +3)}\, \delta_{\ell +1} (z)
 \hat x_{iL}\partial_t \overline\Sigma_i\nonumber \\
 &&\qquad\qquad + {2(2\ell +1)\over c^4(\ell+1)(\ell+2)(2\ell+5)}
 \delta_{\ell+2}(z) \hat x_{ijL} \partial^2_t \overline\Sigma_{ij} \biggr]
 ({\bf x}, t+z |{\bf x}|/c)\ , \label{eq:3.22a}\\
  &&  \nonumber \\
 J_L(t) &=& {\rm FP}_{B=0}\int d^3 {\bf x}|{\bf x}|^B
 \int^1_{-1} dz \biggl[ \delta_\ell (z) \varepsilon_{ab<i_\ell}
  \hat x_{L-1>a}\overline\Sigma_b  \nonumber\\
 &&\qquad\qquad - {2\ell +1\over c^2(\ell+2)(2\ell+3)}
 \delta_{\ell+1}(z) \varepsilon_{ab<i_\ell}\hat x_{L-1>ac} \partial_t
   \overline\Sigma_{bc} \biggr]({\bf x}, t+z |{\bf x}|/c)\ . \label{eq:3.22b}
\end{eqnarray}
\end{subequations}
We have posed
\begin{subequations}
\label{eq:3.23}
\begin{eqnarray}
 \overline\Sigma &=&{\overline\tau^{00} +\overline\tau^{ii}\over c^2}\ ,
 \label{eq:3.23a}\\
 \overline\Sigma_i &=&{\overline\tau^{0i}\over c}\ ,  \label{eq:3.23b}\\
 \overline\Sigma_{ij} &=&\overline\tau^{ij}\ .  \label{eq:3.23c}
\end{eqnarray}
\end{subequations}
The equation (\ref{eq:3.21}) can be solved uniquely for the multipole
moments $M_L$ and $S_L$.  To do so it suffices to notice that for
any gauge term $\partial\xi^{\mu\nu}\equiv \partial^\mu\xi^\nu +\partial^\nu
\xi^\mu - \eta^{\mu\nu}\partial_\lambda\xi^\lambda$ the identity
\begin{equation}
 {1\over 2}\,\partial^2_{ij} [\partial\xi^{00} +\partial\xi^{kk}] +
 \partial_0[\partial_i\partial\xi^{0j}+\partial_j\partial\xi^{0i}] +
 \partial^2_0[\partial\xi^{ij}+{1\over 2}\delta^{ij} (\partial\xi^{00}
  -\partial\xi^{kk})] \equiv 0\  \label{eq:3.24}
\end{equation}
holds. This is nothing but the vanishing of the $0i0j$ component of the
linearized Riemann tensor when computed with $g^{\rm gauge}_{\mu\nu}
=\partial_\mu \xi_\nu +\partial_\nu\xi_\mu$.  Applying (\ref{eq:3.24})
to the gauge term of (\ref{eq:3.21}) and using the form of the multipole
moment decomposition of $h^{\mu\nu}_{\rm can(1)}$ one finds
\begin{subequations}
\label{eq:3.25}
\begin{eqnarray}
 M_L &=& I_L + \delta I_L + O(6)\ , \label{eq:3.25a} \\
 S_L &=& J_L + O(4)\ , \label{eq:3.25b}
\end{eqnarray}
\end{subequations}
where $\delta I_L$ in the mass moments is of 2.5PN order and comes from
the decomposition of $X^{00}$ (which we recall is a retarded solution of
the wave equation) into multipole moments according to
\begin{equation}
 X^{00} = {4G\over c^2} \sum^{+\infty}_{\ell=0} {(-)^\ell\over \ell !}
\partial_L \left[ {1\over r} \delta I_L (t-r/c) \right]\ . \label{eq:3.26}
\end{equation}
There is no contribution coming from $G^2q^{\mu\nu}_{\rm can(2)}$ in the left
side of (\ref{eq:3.21}) thanks to the result proved in Eq.~(\ref{eq:2.19})
in the case where we are using a mass-centred frame. The result
(\ref{eq:3.25a}) generalizes to 2.5PN order the result (4.7a) of
paper~I.  In a future work we shall investigate more systematically the
relations linking the exterior moments $M_L$ and $S_L$ to the source
moments $I_L$ and $J_L$ as they have been defined here.

\section{Explicit expressions of the multipole moments}
\label{sec:4}

Two things must be done in order to obtain the expressions of the
moments $M_L$ and $S_L$ of Eqs.~(\ref{eq:3.25}).  Firstly one
must expand when $c\to +\infty$ the source moments $I_L$ and
$J_L$ of Eqs.~(\ref{eq:3.22}) up to consistent order.
Secondly one must evaluate the 2.5PN modification $\delta
I_L$ entering the mass moments (\ref{eq:3.25a}).

\subsection{The source multipole moments}
\label{sec:4.1}

The post-Newtonian expansion of the source moments (\ref{eq:3.22}) is
straighforwardly performed using a formula which was given in Eq.
(B.14) of Ref.~\cite{BD89}, namely
\begin{eqnarray}
 \int^1_{-1} dz\, \delta_\ell (z) \overline\Sigma ( {\bf x},t+z
   |{\bf x}|/ c) &=&\overline\Sigma ({\bf x},t) +
 {{\bf x}^2\over 2c^2 (2\ell+3)}\,\partial^2_t \overline\Sigma ({\bf x},t)
 \nonumber \\
  &&\qquad + {{\bf x}^4\over 8c^4 (2\ell+3)(2\ell+5)}\,
    \partial^4_t\, \overline\Sigma ({\bf x},t) + O(6)\ . \label{eq:4.1}
\end{eqnarray}
Explicit expressions of the densities $\overline\Sigma$,
$\overline\Sigma_i$ and $\overline\Sigma_{ij}$ have also to be inserted
into $I_L$ and $J_L$; these are easily evaluated with the
help of Eqs.~(\ref{eq:2.8}), (\ref{eq:2.9}), (\ref{eq:3.19}) and
(\ref{eq:3.23}) with the result (identical to paper~I)
\begin{subequations}
\label{eq:4.2}
\begin{eqnarray}
 \overline\Sigma &=& \left[ 1+{4V\over c^2} -{8\over c^4} (W_{ii}-V^2)\right]
 \sigma - {1\over \pi Gc^2}\,\partial_i V \partial_i V \nonumber \\
 && + {1\over \pi Gc^4} \biggl\{ -V \partial^2_t V- 2V_i\partial_t
   \partial_i V - W_{ij}\partial^2_{ij} V-{1\over 2} (\partial_t V)^2
    \nonumber \\
 && \qquad +2 \partial_i V_j \partial_j V_i + 2\partial_i V\partial_i W_{jj}
  - {7\over 2} V\partial_i V\partial_i V \biggr\}\ , \label{eq:4.2a}\\
 \overline\Sigma_i &=& \left[ 1 +{4V\over c^2}\right] \sigma_i  +
  {1\over \pi Gc^2} \left\{ \partial_k V(\partial_i V_k -\partial_k V_i)
  + {3\over 4} \partial_t V \partial_i V \right\} + O(4)\ ,\label{eq:4.2b}\\
 \overline\Sigma_{ij} &=& \sigma_{ij} + {1\over 4\pi G} \left\{ \partial_i
 V \partial_j V - {1\over 2} \delta_{ij} \partial_k V \partial_k V\right\}
  + O(2)\ . \label{eq:4.2c}
\end{eqnarray}
\end{subequations}
Note the important fact that the remainder in Eq.~(\ref{eq:4.1}) is $O(6)$
and not $O(5)$, and thus does not contribute to the 2.5PN order (similarly
the other remainders in (\ref{eq:4.2}) will not contribute). With
Eqs.~(\ref{eq:4.1}) and (\ref{eq:4.2}) we recover the same expressions as
in Eqs.~(4.12) and (4.13) of paper~I for the source moments in
raw form~:
\begin{subequations}
\label{eq:4.3}
\begin{eqnarray}
 I_L(t) &=& {\rm FP}_{B=0}\int d^3 {\bf x}|{\bf x}|^B
 \biggl\{ \biggl[1 + {4\over c^2}V-{8\over c^4} (W_{ii}-V^2)\biggr] \hat x_L
 \sigma - {1\over \pi Gc^2} \hat x_L \partial_i V \partial_i V\nonumber \\
 && + {1\over \pi Gc^4}\hat x_L \biggl[ -V \partial^2_t V- 2V_i\partial_t
   \partial_i V - W_{ij}\partial^2_{ij} V-{1\over 2} (\partial_t V)^2
    \nonumber \\
 &&\qquad\quad\qquad +2 \partial_iV_j\partial_j V_i+2\partial_iV\partial_i
   W_{jj} - {7\over 2} V\partial_i V\partial_i V \biggr]\  \nonumber\\
 && +{|{\bf x}|^2\hat x_L\over 2c^2(2\ell+3)} \partial^2_t
   \left[ \left( 1+{4V\over c^2}\right) \sigma - {1\over \pi Gc^2}
   \partial_i V\partial_i V \right] \nonumber \\
 && +{|{\bf x}|^4\hat x_L\over 8c^4(2\ell+3)(2\ell+5)} \partial^4_t\sigma
   - {2(2\ell+1)|{\bf x}|^2\hat x_{iL}\over c^4(\ell+1)(2\ell+3)(2\ell+5)}
   \partial^3_t\sigma_i  \nonumber\\
 && -{4(2\ell+1)\hat x_{iL}\over c^2(\ell+1)(2\ell+3)}    \partial_t
   \left[ \left( 1 +{4V\over c^2} \right) \sigma_i  +
  {1\over \pi Gc^2} \left\{ \partial_k V(\partial_i V_k -\partial_k V_i)
  + {3\over 4} \partial_t V \partial_i V \right\} \right] \nonumber \\
 && + {2(2\ell+1)\hat x_{ijL}\over c^4(\ell+1)(\ell+2)(2\ell+5)}
    \partial_t^2 \left[ \sigma_{ij} + {1\over 4\pi G} \partial_i V
    \partial_j V \right] \biggr\} + O(6)\ , \label{eq:4.3a} \\
 && \nonumber \\ \nonumber\\
 J_L(t) &=& {\rm FP}_{B=0}\,\varepsilon_{ab<i_\ell}
   \int d^3 {\bf x}|{\bf x}|^B
   \biggl\{ \hat x_{L-1>a}\left( 1+{4\over c^2}V\right) \sigma_b
  + {|{\bf x}|^2\hat x_{L-1>a}\over
   2c^2 (2\ell+3)} \partial^2_t \sigma_b \nonumber \\
 && \qquad\qquad + {1\over \pi Gc^2} \hat x_{L-1>a}
  \left[ \partial_k V(\partial_b V_k -\partial_k V_b)
  + {3\over 4} \partial_t V \partial_b V \right] \nonumber \\
 &&\qquad\qquad - {(2\ell +1)\hat x_{L-1>ac}\over c^2(\ell+2)(2\ell+3)}
 \partial_{t}\left[ \sigma_{bc} +{1\over 4\pi G}  \partial_b V\partial_c
   V \right] \biggr\} + O(4)\ .  \label{eq:4.3b}
\end{eqnarray}
\end{subequations}
The remainders $O(6)$ and $O(4)$ are negligible.
The retarded potentials $V$, $V_i$, $W_{ij}$ are then replaced by their
post-Newtonian expansions when $c\to +\infty$. It is easily seen
that the accuracy of the expansions of $V$ and $W_{ij}$ given in
paper~I is not sufficient and has to be pushed one order
farther.  The relevant expansions are
\begin{subequations}
\label{eq:4.4}
\begin{eqnarray}
 V &=& U + {1\over 2c^2} \partial^2_t X - {2G\over 3c^3} K^{(3)} +
  O(4) \ ,\label{eq:4.4a} \\
 V_i &=& U_i + O(2)\ , \label{eq:4.4b} \\
 W_{ij} &=& P_{ij} - {G\over 2c} \left[ Q^{(3)}_{ij} + {1\over 3}
  \delta_{ij} K^{(3)} \right] + O(2)\ , \label{eq:4.4c}
\end{eqnarray}
\end{subequations}
where the Newtonian-like potentials $U$, $X$, $U_i$ and $P_{ij}$ are
defined as in paper~I by
\begin{subequations}
\label{eq:4.5}
\begin{eqnarray}
 U({\bf x},t) &=& G \int {d^3{\bf x}'\over |{\bf x}-{\bf x}'|} \sigma
 ({\bf x}',t)\ , \label{eq:4.5a}\\
 X({\bf x},t) &=& G \int d^3{\bf x}'|{\bf x}-{\bf x}'| \sigma
 ({\bf x}',t)\ , \label{eq:4.5b}\\
 U_i({\bf x},t) &=& G \int {d^3{\bf x}'\over |{\bf x}-{\bf x}'|} \sigma_i
 ({\bf x}',t)\ , \label{eq:4.5c}\\
 P_{ij}({\bf x},t) &=& G \int {d^3{\bf x}'\over |{\bf x}-{\bf x}'|}
  \left[ \sigma_{ij} + {1\over 4\pi G} \left( \partial_i U\partial_j U
 - {1\over 2} \delta_{ij} \partial_k U\partial_k U\right) \right]
  ({\bf x}',t)\ , \label{eq:4.5d}
\end{eqnarray}
\end{subequations}
and where the new terms involve the tracefree
quadrupole moment $Q_{ij}$ and moment of inertia $K$ associated with the
mass distribution $\sigma$, namely
\begin{subequations}
\label{eq:4.6}
\begin{eqnarray}
 Q_{ij} (t) &=& \int d^3 {\bf x}\,\sigma ({\bf x},t)\,\hat x_{ij} \ ,
     \label{eq:4.6a}\\
 K (t) &=& \int d^3 {\bf x}\,\sigma ({\bf x},t)\,{\bf x}^2\ .\label{eq:4.6b}
\end{eqnarray}
\end{subequations}
When substituting the expansions (\ref{eq:4.4}) into the source moments
(\ref{eq:4.3}) all the terms coming from the Newtonian-like potentials
$U$, $X$, $U_i$ and $P_{ij}$ lead to the same expressions as in paper~I,
while the terms coming from the moments $Q_{ij}$ and $K$ lead to some
correction terms (in $I_L$ only).  Let us write
\begin{subequations}    
\label{eq:4.7}
\begin{eqnarray}
 I_L &=& {\widetilde I}_L +\delta {\widetilde I}_L +O(6)\ ,\label{eq:4.7a} \\
 J_L &=& {\widetilde J}_L +O(4)\ ,  \label{eq:4.7b}
\end{eqnarray}
\end{subequations}
where ${\widetilde I}_L$ and ${\widetilde J}_L$ are the 2PN-accurate moments
which were obtained in paper~I, and where $\delta {\widetilde I}_L$ denotes
a 2.5PN correction term (which is distinct from $\delta I_L$ found in
Eq.~(\ref{eq:3.25a})). After the transformation of ${\widetilde I}_L$ as
in section IV.B of paper~I one can write ${\widetilde I}_L$ and
${\widetilde J}_L$ in the form
\begin{subequations}
\label{eq:4.8}
\begin{eqnarray}
 {\widetilde I}_L(t) &=& {\rm FP}_{B=0}\int d^3 {\bf x}|{\bf x}|^B
 \biggl\{ \hat x_L \biggl[\sigma + {4\over c^4}(\sigma_{ii}U -\sigma P_{ii})
 \biggr] + {|{\bf x}|^2\hat x_L\over 2c^2(2\ell+3)} \partial^2_t
\sigma\nonumber\\
 && -{4(2\ell+1)\hat x_{iL}\over c^2(\ell+1)(2\ell+3)} \partial_t
   \left[ \left( 1 +{4U\over c^2} \right) \sigma_i  +
  {1\over \pi Gc^2} \left( \partial_k U[\partial_i U_k -\partial_k U_i]
  + {3\over 4} \partial_t U \partial_i U \right) \right] \nonumber \\
 && +{|{\bf x}|^4\hat x_L\over 8c^4(2\ell+3)(2\ell+5)} \partial^4_t\sigma
   - {2(2\ell+1)|{\bf x}|^2\hat x_{iL}\over c^4(\ell+1)(2\ell+3)(2\ell+5)}
   \partial^3_t\sigma_i  \nonumber\\
 && + {2(2\ell+1)\over c^4(\ell+1)(\ell+2)(2\ell+5)} \hat x_{ijL}
   \partial_t^2 \left[ \sigma_{ij} + {1\over 4\pi G} \partial_i U
   \partial_j U \right] \nonumber\\
 &&  + {1\over \pi Gc^4}\hat x_L \biggl[ -P_{ij} \partial_{ij}^2 U
  - 2 U_i\partial_t\partial_i U + 2\partial_i U_j \partial_j U_i
  - {3\over 2} (\partial_t U)^2 -U\partial_t^2 U\biggr] \biggr\}\ ,
\label{eq:4.8a}  \\      \nonumber \\
 {\widetilde J}_L(t) &=& {\rm FP}_{B=0}\,\varepsilon_{ab<i_\ell}
   \int d^3 {\bf x}|{\bf x}|^B
   \biggl\{ \hat x_{L-1>a}\left( 1+{4\over c^2}U\right) \sigma_b
  + {|{\bf x}|^2\hat x_{L-1>a}\over
   2c^2 (2\ell+3)} \partial^2_t \sigma_b \nonumber \\
 && \qquad\qquad + {1\over \pi Gc^2} \hat x_{L-1>a}
  \left[ \partial_k U(\partial_b U_k -\partial_k U_b)
  + {3\over 4} \partial_t U \partial_b U \right] \nonumber \\
 &&\qquad\qquad - {(2\ell +1)\hat x_{L-1>ac}\over c^2(\ell+2)(2\ell+3)}
 \partial_{t}\left[ \sigma_{bc} +{1\over 4\pi G}  \partial_b U\partial_c
   U \right] \biggr\}\ .  \label{eq:4.8b}
\end{eqnarray}
\end{subequations}
The moments ${\widetilde I}_L$ and ${\widetilde J}_L$ constituted the
central result of paper~I and they were at the basis of the application
to inspiralling compact binaries in paper~II.  On the other hand the 2.5PN
correction term $\delta {\widetilde I}_L$ is obtained by simple
inspection of Eq.~(\ref{eq:4.3a}).  A simplifying fact in obtaining
$\delta \widetilde I_L$ is that the moments $Q_{ij}$ and $K$ are only
functions of time so that their spatial gradients vanish.  We obtain
\begin{equation}
 \delta {\widetilde I}_L = {\rm FP}_{B=0} \int d^3 {\bf x}
  |{\bf x}|^B \biggl\{{2G\over 3c^5} K^{(3)} \sigma \hat x_L
  + {1\over 2\pi c^5} Q_{ij}^{(3)} \hat x_L \partial^2_{ij} U\biggr\}\ .
      \label{eq:4.9}
\end{equation}
The second term is an integral having a priori a non-compact support.
However it can be transformed into a manifestly compact-support form by
means of the formula
\begin{equation}
{\rm FP}_{B=0} \int d^3 {\bf x} |{\bf x}|^B {\hat x_L\over |{\bf x}-{\bf y}|}
 = - {2\pi\over 2\ell+3} |{\bf y}|^2 \hat y_L\ . \label{eq:4.10}
\end{equation}
This formula is proved by noticing that the integral defined by $I_B= \int
d^3{\bf x}|{\bf x}|^B \hat x_L |{\bf x}-{\bf y}|^{-1}$ is proportional to
a prefactor $|{\bf y}|^B$ and satisfies $\Delta_{\bf y} I_B
= -4\pi |{\bf y}|^B \hat y_L$.  These two facts imply that $I_B =
\Delta_{\bf y}^{-1} (-4\pi |{\bf y}|^B \hat y_L)$ where $\Delta_{\bf
y}^{-1}$ is defined as in Eq. (3.9) of Ref. \cite{BD86}, and thus that $I_B
= -4\pi |{\bf y}|^{B+2}\hat y_L /(B+2)(B+2\ell+3)$ which yields
(\ref{eq:4.10}) after taking the finite part. The more complicated
formula (4.23) in paper~I can interestingly be compared with
(\ref{eq:4.10}). Thanks to (\ref{eq:4.10}) we can write
\begin{equation}
 \delta {\widetilde I}_L = {G\over c^5} \int d^3 {\bf x} \left\{
{2\over 3}\, K^{(3)} \sigma \hat x_L - {1\over 2\ell+3} Q^{(3)}_{ij} \sigma
 \partial^2_{ij} [|{\bf x}|^2 \hat x_L] \right\}\ . \label{eq:4.11}
\end{equation}
Expanding the spatial derivatives in the second term yields finally
\begin{equation}
 \delta {\widetilde I}_L = {G\over c^5} \left\{ {2\over 3}\, K^{(3)} Q_L
  - {4\ell\over 2\ell+3} Q^{(3)}_{k<i_\ell} Q_{L-1>k}
  - {\ell(\ell-1)\over 2\ell-1} Q^{(3)}_{<i_\ell i_{\ell-1}}
     K_{L-2>} \right\} \label{eq:4.12}
\end{equation}
where we have posed
\begin{subequations}
\label{eq:4.13}
\begin{eqnarray}
 Q_L(t) &=& \int d^3 {\bf x}\, \sigma\, \hat x_L\ , \label{eq:4.13a} \\
 K_{L-2}(t) &=& \int d^3 {\bf x}\, \sigma\, {\bf x}^2\,\hat x_{L-2}\ .
     \label{eq:4.13b}
\end{eqnarray}
\end{subequations}
These definitions are in conformity with the earlier definitions
(\ref{eq:4.6}). [The brackets $<>$ denote the STF projection.]

\subsection{The 2.5PN modification of the mass moments}
\label{sec:4.2}

In addition to the previous contribution $\delta {\widetilde I}_L$ which is
part of the source multipole moments $I_L$, we have seen in
Eq.~(\ref{eq:3.25a}) that there exists also a 2.5PN contribution $\delta
I_L$ entering $M_L$. Evidently the contribution $\delta I_L$ is as important
as $\delta {\widetilde I}_L$ in that it contributes also to the asymptotic
waveform depending on the moments $M_L$. [Note that the terminology
refering to $I_L$ and $J_L$ as {\it the} multipole moments of the source
by contrast to $M_L$ and $S_L$ which are viewed as some intermediate
moments devoided of direct physical signification is somewhat arbitrary.
What only matters {\it in fine} is to express (by any convenient means)
the asymptotic waveform in terms of physical quantities belonging to the
source.]

The 2.5PN term $\delta I_L$ results from the multipole decomposition
of the quantity $X^{00}$ which is itself determined from the other quantity
$\Omega^{00}$. See Eqs.~(\ref{eq:3.26}), (\ref{eq:3.14a}) and (\ref{eq:3.11a}).
The first step in the computation of $\delta I_L$ is to find the vector
$\varphi^\mu$ of the coordinate transformation between the inner and
external metrics.  Since $\varphi^\mu$ is of order $O(3,4)$ and is
d'Alembertian-free to order $ O(7,8)$ [see Eqs.~(\ref{eq:3.2}) and
(\ref{eq:3.3})], there exist four sets of STF tensors $W_L(t)$,
$X_L(t)$, $Y_L(t)$ and $Z_L(t)$ such that
\begin{subequations}
\label{eq:4.14}
\begin{eqnarray}
 \varphi^0 &=&-{4G\over c^3} \sum_{\ell\geq 0}{(-)^\ell\over\ell!}\partial_L
   \left[ {1\over r} W_L (t-r/c) \right] + O(7)\ , \label{eq:4.14a} \\
 \varphi^i &=&{4G\over c^4} \sum_{\ell\geq 0}{(-)^\ell\over\ell!}\partial_{iL}
   \left[ {1\over r} X_L (t-r/c) \right] \nonumber \\
  &&+{4G\over c^4} \sum_{\ell\geq 1}{(-)^\ell\over\ell!} \left\{\partial_{L-1}
   \left[ {1\over r} Y_{iL-1}(t-r/c) \right]\right. \nonumber\\
  &&\qquad \quad\left. +{\ell\over \ell+1}
   \varepsilon_{iab} \partial_{aL-1} \left[{1\over r} Z_{bL-1}(t-r/c)\right]
   \right\} + O(8)\ . \label{eq:4.14b}
\end{eqnarray}
\end{subequations}
The powers of $1/c$ in front of these terms are such that $W_L,\cdots,
Z_L$ have a non-zero limit when $c\to +\infty$.  We compute these
tensors to the lowest order in $1/c$.  To do this let us recall the
relations between $V^{\rm ext}_i$ and $W^{\rm ext}_{ij}$ and the
multipole expansions ${\cal M}(V_i)$ and ${\cal M}(W_{ij})$ as obtained
in Eqs.~(\ref{eq:3.6b}) and (\ref{eq:3.6c}).  We have
\begin{subequations}
\label{eq:4.15}
\begin{eqnarray}
 V^{\rm ext}_i &=& G \sum_{\ell\geq 0} {(-)^\ell\over \ell!} \partial_L
   \left[ {1\over r} {\cal V}_i^L (t-r/c) \right] - {c^3\over 4} \partial_i
  \varphi^0 + O(2)\ , \label{eq:4.15a} \\
 V^{\rm ext}_{ij} &=& G \sum_{\ell\geq 0} {(-)^\ell\over \ell!} \partial_L
   \left[ {1\over r} {\cal W}_{ij}^L (t-r/c) \right] - {c^4\over 4}
  [\partial_i \varphi^j + \partial_j\varphi^i -\delta_{ij} (\partial_0
  \varphi^0 + \partial_k\varphi^k)] + O(2)\ , \label{eq:4.15b}
\end{eqnarray}
\end{subequations}
where we have transformed the relation for $W^{\rm ext}_{ij}$ into a
simpler relation for $V^{\rm ext}_{ij}$, and
where the moments take to lowest order the form [see
Eqs.~(\ref{eq:3.8b}) and (\ref{eq:3.8c})]~:
\begin{subequations}
\label{eq:4.16}
\begin{eqnarray}
 {\cal V}^L_i (t) &=& \int d^3 {\bf x}\, \hat x_L \sigma_i ({\bf x},t)
   + O(2)\ , \label{eq:4.16a} \\
 {\cal W}^L_{ij} (t) &=& {\rm FP}_{B=0} \int d^3 {\bf x} |{\bf x}|^B
   \hat x_L\,\overline\Sigma_{ij} ({\bf x},t) + O(2)\ . \label{eq:4.16b}
\end{eqnarray}
\end{subequations}
The (non-compact-supported) stress density $\overline\Sigma_{ij}$ is defined
by Eq.~(\ref{eq:4.2c}). Having written (\ref{eq:4.15})-(\ref{eq:4.16}) and
knowing the explicit multipole decompositions of $V^{\rm ext}_i$ and
$V^{\rm ext}_{ij}$ given by Eqs.~(\ref{eq:2.11b}) and (\ref{eq:2.11c}),
it is a simple matter to
compute the tensors $W_L$,\dots,$Z_L$ by decomposition of the
integrands $\hat x_L \sigma_i$ and $\hat x_L\,\overline\Sigma_{ij}$ entering
(\ref{eq:4.16}) into irreducible tensorial pieces with respect to their
$\ell +1$ and $\ell +2$ indices.  We do not detail this computation but
simply give the result, which is
\begin{subequations}
\label{eq:4.17}
\begin{eqnarray}
 W_L &=& {2\ell+1\over (\ell+1)(2\ell+3)} \int d^3{\bf x}\,\hat x_{iL}
  \sigma_i + O(2)\ , \label{eq:4.17a}\\
 X_L &=& {2\ell+1\over 2(\ell+1)(\ell+2)(2\ell+5)} {\rm FP}_{B=0}
   \int d^3{\bf x} |{\bf x}|^B \hat x_{ijL} \overline\Sigma_{ij}
    + O(2)\ , \label{eq:4.17b}\\
Y_L &=& {3\ell(2\ell-1)\over (\ell+1)(2\ell+3)}{\rm FP}_{B=0} \int d^3{\bf x}
  |{\bf x}|^B \left(\hat x_{i<L-1} \overline\Sigma_{i_\ell>i}
 -{1\over 3} \hat x_L \overline\Sigma_{ii} \right) + O(2)\ ,\label{eq:4.17c}\\
 Z_L &=&-{2\ell+1\over (\ell+2)(2\ell+3)} {\rm FP}_{B=0}
   \int d^3{\bf x} |{\bf x}|^B \varepsilon_{ab<i_\ell} \hat x_{L-1>bc}
    \overline\Sigma_{ac}     + O(2)\ . \label{eq:4.17d}
\end{eqnarray}
The tensor $W_L$ is manifestly of compact-supported form.
There is agreement for this tensor with the previous result obtained in
Eqs. (2.22a) and (2.19c) of Ref. \cite{BD89}.  As they are written the
other tensors $X_L$, $Y_L$, $Z_L$ do not have a compact-supported form.
However $Y_L$ can be re-written equivalently in such a form:
\begin{equation}
 Y_L =  \int d^3 {\bf x} \left\{ {3(2\ell+1)\over (\ell+1)(2\ell+3)}
 \hat x_{iL} \partial_t \sigma_i - \hat x_L \left( \sigma_{ii}
 - {1 \over 2} \sigma U \right) \right\} +O(2)\ . \label{eq:4.17e}
\end{equation}
\end{subequations}
The transformation of $Y_L$ into the form
(\ref{eq:4.17e}) is done using the results (4.2) and (4.18) of paper~I.

The quantities $\Omega^{00}$ and $X^{00}$ can now be evaluated.  As is
clear from its structure and the form of $\varphi^\mu$ and ${\cal M} (V)$,
${\cal M} (V_i)$, the quantity $\Omega^ {00}$ is made up of a sum of
quadratic products of retarded waves.  We shall write
\begin{equation}
\Omega^{00} = {1\over c^6} \sum_{p,q} \hat\partial_P \left[ {1\over r}
 F(t-r/c) \right] \hat\partial_Q \left[ {1\over r} G(t-r/c) \right]
 + O(10) \label{eq:4.18}
\end{equation}
where $F$ and $G$ are some functions of the retarded time symbolizing some
derivatives of the functions ${\cal V}^L$, ${\cal V}^L_i$ and $W_L$, $X_L$,
$Y_L$, $Z_L$ (all indices suppressed). We assume (as can always be done)
that the derivative operators are trace-free: $\hat\partial_P \equiv
\partial_{<i_1} \partial_{i_2}\cdots \partial_{i_p>}$ and $\hat\partial_Q
\equiv \partial_{<j_1} \partial_{j_2}\cdots \partial_{j_q>}$. The power of
$1/c$ in front indicates the true order of magnitude of $\Omega^{00}$ when
$c\to +\infty$ [see (3.12)], and the remainder $O(10)$ comes from
the un-controlled remainder terms in (\ref{eq:4.14}). To evaluate $X^{00}$
we need to know the action of the operator ${\rm FP}_{B=0} \Box^{-1}_R
r^B \Box -\openone$, where $\openone$ denotes the unit operator, on the
generic term composing $\Omega^{00}$. Actually we shall be interested only
in that part of $X^{00}$ which is strictly larger than the remainder $O(8)$
we neglect in (\ref{eq:3.16}). In that case, a useful formula shows that
all the terms in $\Omega^{00}$ which are composed of the product of two
waves with multipolarities $p\geq 1$ {\it and} $q\geq 1$ yield negligible
terms in $X^{00}$.  This formula, which is proved in Appendix A, reads
\begin{eqnarray}
&&({\rm FP}_{B=0} \Box^{-1}_R r^B \Box -\openone) \left\{ \hat\partial_P
  \left[{1\over r} F(t-r/c)\right] \hat\partial_Q
  \left[{1\over r} G(t-r/c)\right] \right\}\nonumber\\
&&\qquad\qquad\qquad =  {1\over c}\hat\partial_{PQ}
 \left[ {1\over r} \biggl(\delta_{p,0} F^{(1)} G
  + \delta_{0,q} FG^{(1)} \biggr) \right] + O(3)\ \hfill \label{eq:4.19}
\end{eqnarray}
($\delta_{p,q}$ is the Kronecker symbol). When both $p\geq 1$ and $q\geq 1$
the right side of (\ref{eq:4.19}) is of order $O(3)$ relatively to the left
side and the corresponding term in $X^{00}$ is of negligible order $O(9)$.
We can thus limit our consideration to the terms in $ \Omega^{00}$ involving
at least one monopolar wave $p=0$ or $q=0$. We insert into Eq.~(\ref{eq:3.11a})
the multipole expansions (\ref{eq:3.7a}) and (\ref{eq:3.7b}) together with
those of $\varphi^0, \varphi^i$ given by (\ref{eq:4.14}). By straightforward
application of (\ref{eq:4.19}) to each of the resulting terms one finds
\begin{equation}
 X^{00} = {16G^2\over c^7} \sum_{\ell\geq 0} {(-)^\ell\over \ell !}
 \partial_L \left[ {1\over r} \left( W^{(2)} {\cal V}_L
  - W^{(1)} {\cal V}^{(1)}_L + \ell\,Y^{(1)}_{<i_\ell}
   {\cal V}_{L-1>} \right) \right] + O(9)\ , \label{eq:4.20}
\end{equation}
where the functions $W(t)$ and $Y_i(t)$ are given by (\ref{eq:4.17a}) in
which $\ell=0$ and by (\ref{eq:4.17c}) in which $\ell=1$, and where we have
used the law of conservation of mass implying ${\cal V}^{(1)} =O(2)$ and our
assumption of mass-centered frame implying ${\cal V}^{(1)}_i =O(2)$. [See
the definitions of the functions ${\cal V}^L$ and ${\cal V}^L_i$ in
Eqs.~(\ref{eq:3.8a}) and (\ref{eq:3.8b})~;  in (\ref{eq:4.20}) we
denote ${\cal V}_L \equiv {\cal V}^L$ and ${\cal V}_{L-1} \equiv {\cal
V}^{L-1}$ for the function (\ref{eq:3.8a}) although this notation is slightly
ambiguous with (\ref{eq:3.8b}).] To lowest order the function
${\cal V}_L(t)$ reduces to
\begin{equation}
 {\cal V}_L = Q_L + O(2)\ , \label{eq:4.21}
\end{equation}
where $Q_L(t)$ is the moment defined in Eq.  (\ref{eq:4.13a}).  On the
other hand $W(t)$ satisfies
\begin{equation}
 W = {1\over 3} \int d^3 {\bf x}\, x_i \sigma_i + O(2) = {1\over 6}\,K^{(1)}
  + O(2)\ ,  \label{eq:4.22}
\end{equation}
where $K(t)$ is the moment of inertia (\ref{eq:4.6b}). Similarly one finds
\begin{subequations}
\label{eq:4.23}
\begin{equation}
   Y_i = {1\over 5}\, G^{(1)}_i + O(2) \label{eq:4.23a}
\end{equation}
where the vector $G_i(t)$ reads
\begin{equation}
   G_i = \int d^3 {\bf x} \left(\sigma_i {\bf x}^2-{1\over 2}
    \sigma_j x_i x_j \right)   \ . \label{eq:4.23b}
\end{equation}
\end{subequations}
With these notations we end up with the 2.5PN correction term $\delta I_L$
[compare (\ref{eq:3.26}) and (\ref{eq:4.20})],
\begin{equation}
  \delta I_L = {G\over c^5} \left\{ {2\over 3}\, K^{(3)} Q_L - {2\over 3}
 K^{(2)} Q^{(1)}_L + {4\ell\over 5} G^{(2)}_{<i_\ell} Q_{L-1>}\right\}
  + O(7)\ . \label{eq:4.24}
\end{equation}

Summarizing the results so far, we have explicitly computed the mass
multipole moment $M_L$ given by Eq.  (\ref{eq:3.25a}).  It contains a
2.5PN contribution issued from the source moment $I_L$ and computed in
Eq.~(\ref{eq:4.12}), and also the direct 2.5PN modification computed in
Eq.~(\ref{eq:4.24}).  We can write
\begin{equation}
 M_L = {\widetilde I}_L + \Delta I_L + O(6) \label{eq:4.25}
\end{equation}
where ${\widetilde I}_L$ is given by Eq.~(\ref{eq:4.8a}) (this was the
result of paper~I), and where $\Delta I_L = \delta {\widetilde I}_L +
\delta I_L$ is given by
\begin{eqnarray}
 \Delta I_L &=& {G\over c^5} \biggl\{ {4\over 3} K^{(3)} Q_L
  -{2\over 3} K^{(2)} Q^{(1)}_L - {\ell(\ell-1)\over 2\ell -1} K_{<L-2}
  Q^{(3)}_{i_{\ell-1} i_\ell >} \nonumber \\
 && \qquad \qquad
  -{4\ell\over {2\ell +3}} Q^{(3)}_{k<i_\ell} Q_{L-1>k} +{4\ell\over 5}
   G^{(2)}_{<i_\ell} Q_{L-1>} \biggr\} + O(7) \ . \label{eq:4.26}
\end{eqnarray}
We recall that the tensors $Q_L$, $K_{L-2}$ and $G_i$ are defined in
Eqs.~(\ref{eq:4.13}) and (\ref{eq:4.23b}). The low orders in $\ell$ read
\begin{subequations}
\label{eq:4.27}
\begin{eqnarray}
 \Delta I  &=& {4G\over 3c^5} MK^{(3)} + O(7)\ , \label{eq:4.27a}  \\
 \Delta I_i &=& {4G\over 5c^5} MG^{(2)}_i + O(7)\ , \label{eq:4.27b} \\
 \Delta I_{ij} &=& {G\over c^5} \left\{ {4\over 3} K^{(3)} Q_{ij}
   -{2\over 3} K^{(2)} Q^{(1)}_{ij} - {2\over 3} KQ^{(3)}_{ij}
   -{8\over 7} Q^{(3)}_{k<i} Q_{j>k} \right\} + O(7)\ , \label{eq:4.27c}
\end{eqnarray}
\end{subequations}
where $M$ is the total mass such that $Q=M+O(2)$, and where we have
used a frame such that $Q_i=O(2)$. The quadrupolar correction
term (\ref{eq:4.27c}) will contribute to the asymptotic waveform at the
2.5PN order. The dipolar correction term (\ref{eq:4.27b}) will be used
to determine the center of mass of the system at this order.

\section{The 2.5PN-accurate gravitational luminosity}  \label{sec:5}

The mass and current multipole moments $M_L$ and $S_L$ are determined up
to the neglect of $O(6)$ and $O(4)$ terms respectively and can be used to
compute the gravitational luminosity (or energy loss rate) of the system
at 2.5PN order. To compute the waveform at the same order would necessitate
a more accurate determination of the current moments, up to the
neglect of $O(5)$ terms. We shall leave this computation for future work.

Let $X^\mu =(cT,{\bf X})$ be a coordinate system valid in a neighbourhood
of future null infinity and such that the metric admits a Bondi-type
expansion when $R\equiv |{\bf X}| \to +\infty$ with $T_R\equiv T-R/c$
staying constant.  See \cite{B} for the proof (within the present
formalism) of the existence and construction of such a coordinate
system.  The relation between $T_R$ and the retarded time of the
harmonic coordinates $x^{\mu}_{\rm can}$ is
\begin{equation}
 T_R = t_{\rm can} - {r_{\rm can}\over c} - {2GM\over c^3}
 \ln \left( {r_{\rm can}\over cb} \right) + O(1/r_{\rm can})
  + O(5)\ , \label{eq:5.1}
\end{equation}
where $b$ is some arbitrary constant time-scale. [Actually, to be consistent
with the 2.5PN precision one should consider also the next-order
post-Newtonian term in Eq.~(\ref{eq:5.1})~;  however this term will not
be needed in the following.] It is sufficient to control the transverse
and tracefree projection of the leading-order term $\sim R^{-1}$ in the
spatial metric.  A multipole decomposition yields a parametrization into
two and only two sets of STF moments $U_L$ and $V_L$ which depend on
$T_R$ and can be referred to as the ``radiative" or ``observable" mass and
current moments. These are chosen so that they reduce in the limit $c\to +
\infty$ to the $\ell$th time derivatives of the ordinary Newtonian
mass and current moments of the source. The total luminosity ${\cal L}
={\cal L}(T_R)$ of the gravitational wave emission when expressed in
terms of $U_L$ and $V_L$ reads \cite{Th80}
\begin{equation}
 {\cal L} = \sum^{+\infty}_{\ell=2} {G\over c^{2\ell+1}} \left\{
   {(\ell+1)(\ell+2)\over (\ell-1)\ell \ell!(2\ell+1)!!} U^{(1)}_L
   U^{(1)}_L + {4\ell (\ell+2)\over (\ell-1)(\ell+1)!(2\ell+1)!!c^2}
   V^{(1)}_L V^{(1)}_L \right\} \ . \label{eq:5.2}
\end{equation}
Considering ${\cal L}$ to 2.5PN order one retains in (\ref{eq:5.2})
all the terms up to the neglect of a remainder $O(6)$, and finds
\begin{eqnarray}
 {\cal L} = {G\over c^5} \biggl\{ {1\over 5} U^{(1)}_{ij} U^{(1)}_{ij}
 &+&{1\over c^2} \left[ {1\over 189} U^{(1)}_{ijk} U^{(1)}_{ijk}
 + {16\over 45} V^{(1)}_{ij} V^{(1)}_{ij}\right] \nonumber \\
 &+&{1\over c^4} \left[ {1\over 9072} U^{(1)}_{ijkm} U^{(1)}_{ijkm}
 + {1\over 84} V^{(1)}_{ijk} V^{(1)}_{ijk}\right] + O(6) \biggr\}\ .
 \label{eq:5.3}
\end{eqnarray}
Because the powers of $1/c$ go by steps of two in ${\cal L}$, this
expression is in fact the same as already used in paper~I.

All the problem is to find the relations between the radiative moments
$U_L$, $V_L$ and the moments $M_L$, $S_L$ we have previously determined.
We rely on previous papers  (\cite{BD89,BD92} and
paper~I) having written the general form of these relations as
\begin{subequations}
\label{eq:5.4}
\begin{eqnarray}
 U_L (T_R) &=& M_L^{(\ell)} (T_R) + \sum_{n\geq 2} {G^{n-1}\over
  c^{3(n-1)+\Sigma \underline{\ell}_i-\ell}} X_{nL} (T_R)\ ,\label{eq:5.4a}\\
\varepsilon_{ai_\ell i_{\ell -1}} V_{aL-2} (T_R) &=&
  \varepsilon_{ai_\ell i_{\ell -1}} S^{(\ell -1)}_{aL-2} (T_R)
   + \sum_{n\geq 2} {G^{n-1}\over
c^{3(n-1)+\Sigma \underline{\ell}_i-\ell}} Y_{nL} (T_R)\ , \label{eq:5.4b}
\end{eqnarray}
\end{subequations}
where the functions $X_{nL}$ and $Y_{nL}$ represent some nonlinear (and in
general non-local) functionals of the moments $M_L$ and $S_L$. The powers
of $1/c$ in Eqs.~(\ref{eq:5.4}) come from the dimensionality of the
functionals $X_{nL}$ and $Y_{nL}$ which is chosen to be that of a product
of $n$ multipole moments and their time derivatives. We can write symbolically
\begin{equation}
  X_{nL},\ Y_{nL} \ \sim  \ M^{(a_1)}_{L_1} M^{(a_2)}_{L_2} \cdots
    S^{(a_n)}_{L_n}\ . \label{eq:5.5}
\end{equation}
The notation in Eqs.~(\ref{eq:5.4})-(\ref{eq:5.5}) is the same as in paper~I~;
in particular $\Sigma \ell_i$ denotes the total number of indices on the
$n$ moments in the term in question, and $\Sigma \underline{\ell}_i$ denotes
the sum $\Sigma \ell_i + s$ where $s$ is the number of current moments.
Here we shall need only the fact that $\Sigma \underline\ell_i$
is larger than the multipolarity $\ell$ by an even positive integer $2k$
which represents the number of contracted indices between the moments
composing the term (with the current moments carrying their associated
Levi-Civita symbol), i.e.
\begin{equation}
  \Sigma \underline \ell_i = \ell + 2k \ . \label{eq:5.6}
\end{equation}
With the latter equation it is simple to control the type of nonlinearities
which are present in the radiative moments to 2.5PN order. Since the
reasoning has already been done in paper~I to 2PN order we consider only
the case which is further needed, that of a term of pure 2.5PN order in the
mass-type quadrupole moment $U_{ij}$ (having $\ell = 2$). By
Eqs.~(\ref{eq:5.4a}) and (\ref{eq:5.6}) this case corresponds to
$3(n-1)+2k =5.$ The only solution is $n=2$ (quadratic nonlinearity) and
$k=1$ (one contraction of indices between the moments). With two moments,
one contraction and $\ell=2$ one has $\underline\ell_1 +
\underline\ell_2 =4$.  Furthermore one of the two moments is non-static,
hence $\underline\ell_2 \geq 2$ say, so we obtain only two possibilities
$(\underline\ell_1$, $\underline\ell_2) =(1,3)$ or (2,2).  The first
possibility is excluded because the moment having $\underline\ell_1 =1$
is necessarily the constant mass dipole $M_i$ which has been set to zero.
Only remains the second possibility
$\underline\ell_1 =\underline\ell_2 =2$ which corresponds either to the
interaction between two mass-type quadrupole moments $M_{ij}$ or to the
interaction of the (constant) current-type dipole $S_i$ with $M_{ij}$.

We combine these facts with the study done in Ref.~\cite{BD92} of the
occurence of ``hereditary'' terms in the asymptotic metric at the
quadratic approximation $n=2$. Two and only two types of hereditary terms
were found:  the ``tail'' terms coming from the interaction between the
monopole $M$ and non-static multipoles, and the nonlinear ``memory'' term
which is made of the interaction between two non-static multipoles. The tail
terms are of order $c^{-3}$ and have been included in paper~I, but the
memory term arises at the order $c^{-5}$ (in the radiative quadrupole
$U_{ij}$).  The latter term can be straightforwardly computed from
Eqs.~(2.42a), (2.21) and (2.11a) in Ref.~\cite{BD92}. An equivalent result
can be found in Ref.~\cite{WiW91}. For discussions on the memory term see
Refs.~\cite{Chr91,Th92}. It is clear by the previous reasoning that the
memory term represents the hereditary part of the 2.5PN contribution
in the radiative moment $U_{ij}$, corresponding to the interaction of
two moments $M_{ij}$. Associated with this term there are also some
instantaneous terms having the same structure and exhausting (a priori)
the possibilities of sharing time-derivatives between the two moments.

Gathering these results with the results of Ref.~\cite{BD92} and paper~I
we obtain the expression of the radiative quadrupole $U_{ij}$ to 2.5PN
order as follows~:
\begin{eqnarray}
 U_{ij} (T_R) &=& M^{(2)}_{ij} (T_R) + {2GM\over c^3} \int^{+\infty}_0 d\tau
  \left[ \ln \left({\tau\over 2b}\right) + {11\over 12}\right] M^{(4)}_{ij}
   (T_R-\tau) \nonumber\\
  && + {G\over c^5} \biggl\{ -{2\over 7} \int^{T_R}_{-\infty} du\,
 M^{(3)}_{k<i} (u) M^{(3)}_{j>k}(u) + \alpha M^{(3)}_{k<i} M^{(2)}_{j>k}
   \nonumber \\
 && \qquad + \beta M^{(4)}_{k<i} M^{(1)}_{j>k}+\gamma M^{(5)}_{k<i}
   M_{j>k} + \lambda S_k M^{(4)}_{m<i}
   \varepsilon_{j>km}\biggr\}+O(6)\ .\label{eq:5.7}
\end{eqnarray}
The constant $b$ in the term of order $c^{-3}$
(tail integral) is the same as in Eq.~(\ref{eq:5.1}).
The memory term is the integral in the brackets of order $c^{-5}$.
The qualitatively different nature of these two integrals can be clearly
understood when taking the limit $T_R\to +\infty$ corresponding to very
late times after the system has ceased to emit radiation. In this limit the
third and higher time derivatives of $M_{ij}$ are expected to tend to
zero, so the tail term tends to zero while by contrast the memory term
tends to the finite limit $-(2G/7c^5) \int^{+\infty}_{-\infty} du
M^{(3)}_{k<i} M^{(3)}_{j>k}$ (see Refs.~\cite{Chr91,Th92}).
The coefficients $\alpha$, $\beta$, $\gamma$ and $\lambda$ are some purely
numerical coefficients in front of instantaneous terms (which depend on
$T_R$ only).  These coefficients can be obtained by a long computation
using the algorithm of Ref.~\cite{BD86}, however we shall not need them
in the application below (they will be computed in a future work).  The
higher-order radiative moments take similar expressions but are needed
only to a lower precision.  The relevant expressions for $U_{ijk}$ and
$V_{ij}$ have been written in paper~I, and read
\begin{subequations}
\label{eq:5.8}
\begin{eqnarray}
 U_{ijk} (T_R) &=& M^{(3)}_{ijk} (T_R) + {2GM\over c^3} \int^{+\infty}_0
  d\tau \left[ \ln \left({\tau\over 2b}\right) + {97\over 60}\right]
  M^{(5)}_{ijk} (T_R-\tau) + O(5)\ , \label{eq:5.8a} \\
 V_{ij} (T_R) &=& S^{(2)}_{ij} (T_R) + {2GM\over c^3} \int^{+\infty}_0 d\tau
  \left[ \ln \left({\tau\over 2b}\right) + {7\over 6}\right] S^{(4)}_{ij}
   (T_R-\tau) + O(5)\ , \label{eq:5.8b}
\end{eqnarray}
\end{subequations}
while the other needed moments are given by
\begin{subequations}
\label{eq:5.9}
\begin{eqnarray}
 U_{ijkm} (T_R) &=& M^{(4)}_{ijkm} (T_R) + O(3)\ , \label{eq:5.9a}\\
 V_{ijk} (T_R) &=& S^{(3)}_{ijk} (T_R) + O(3)\ . \label{eq:5.9b}
\end{eqnarray}
\end{subequations}

The expressions (\ref{eq:5.7})-(\ref{eq:5.9}) of the radiative moments are
to be inserted into the gravitational luminosity (\ref{eq:5.3}).
This leads to a natural (though not unique) decomposition of ${\cal L}$
into instantaneous and tail contributions,
\begin{equation}
  {\cal L} = {\cal L}_{\rm inst} + {\cal L}_{\rm tail}\ . \label{eq:5.10}
\end{equation}
The contribution
${\cal L}_{\rm inst}$ depends only on the instant $T_R$ and is given by
\begin{eqnarray}
 {\cal L}_{\rm inst} &=& {G\over c^5} \biggl\{ {1\over 5} M^{(3)}_{ij}
 M^{(3)}_{ij} + {1\over c^2} \left[ {1\over 189} M^{(4)}_{ijk} M^{(4)}_{ijk}
 + {16\over 45} S^{(3)}_{ij} S^{(3)}_{ij}\right]
 + {1\over c^4} \left[ {1\over 9072} M^{(5)}_{ijkm} M^{(5)}_{ijkm}
 + {1\over 84} S^{(4)}_{ijk} S^{(4)}_{ijk}\right] \nonumber\\
 &&\qquad +{2G\over 5c^5}M^{(3)}_{ij}\left[(\alpha -2/7)M^{(3)}_{ik}
  M^{(3)}_{jk}+(\alpha +\beta) M^{(4)}_{ik} M^{(2)}_{jk}\right. \nonumber\\
 && \qquad\qquad + (\beta + \gamma) M^{(5)}_{ik} M^{(1)}_{jk}
 + \gamma M^{(6)}_{ik} M_{jk} \left.
 + \lambda S_k M^{(5)}_{mi} \varepsilon_{jkm} \right]
  + O(6) \biggr\}\ . \label{eq:5.11}
\end{eqnarray}
Note that this involves the term coming from the nonlinear
memory which is instantaneous in the energy loss (the memory effect exists
only in the waveform). The tail contribution depends
on all instants $T_R-\tau$ anterior to $T_R$ and reads
\begin{eqnarray}
 {\cal L}_{\rm tail}&=& {4G^2M\over c^5}
    \biggl\{ {1\over 5c^3} M^{(3)}_{ij} (T_R) \int^{+\infty}_0 d\tau
  M^{(5)}_{ij} (T_R-\tau) \ln \left( {\tau\over 2b_1}\right)\nonumber\\
 &&\qquad\qquad\qquad
  + {1\over 189c^5} M^{(4)}_{ijk} (T_R) \int^{+\infty}_0 d\tau
  M^{(6)}_{ijk} (T_R-\tau) \ln \left( {\tau\over 2b_2}\right)\nonumber\\
 && \qquad\qquad\qquad
  + {16\over 45c^5} S^{(3)}_{ij} (T_R)  \int^{+\infty}_0 d\tau
  S^{(5)}_{ij} (T_R-\tau) \ln
  \left( {\tau\over 2b_3}\right) +O(6) \biggr\} \ , \label{eq:5.12}
\end{eqnarray}
where we have posed for simplicity
\begin{equation}
 b_1 = b\,e^{-11/12}\ , \quad b_2= b\,e^{-97/60}\ , \quad
 b_3 = b\, e^{-7/6}\ . \label{eq:5.13}
\end{equation}
The luminosity (\ref{eq:5.10})-(\ref{eq:5.13}) in which
the moments have been determined in Sect.~IV is our final result for the
general case of a (semi relatistic) isolated system.

\section{Application to inspiralling compact binaries} \label{sec:6}

The authors of Ref.~\cite{BDI95} (paper~II) applied the results of paper~I
to an inspiralling compact binary system modelled by two point-masses moving
on a circular orbit. Here we do the same for the results derived
previously and obtain the energy loss rate and associated laws of
variation of the frequency and phase of the binary to 2.5PN order.

\subsection{The equations of motion} \label{sec:6.1}

Obviously the equations of motion of two point-masses at the 2.5PN
approximation are needed for this application. These equations have been
obtained in the same coordinates as used here by Damour and Deruelle
\cite{DD} studying the dynamics of the binary pulsar.
For inspiralling compact binaries one needs only to specialize these
equations to the case of an orbit which is circular (apart from the
gradual inspiral).  A relevant summary of the Damour-Deruelle equations
of motion is presented for the reader's convenience in Appendix~B.
Here we quote the results valid for circular orbits, following mostly
the notation of paper~II.  Mass parameters are denoted by
\begin{equation}
  m \equiv m_1 + m_2\ ;\quad X_1 \equiv {m_1\over m}\ ;
   \quad X_2 \equiv {m_2\over m}\ ;
  \quad \nu \equiv X_1X_2 \label{eq:6.1}
\end{equation}
(the total mass is henceforth denoted by $m$ to conform with paper~II).
The individual positions of the two bodies in harmonic coordinates are
$y^i_1$ and $y^i_2$.  Their relative separation and relative velocity are
\begin{equation}
  x^i = y^i_1 -y^i_2\quad ; \quad v^i = {dx^i\over dt}\ . \label{eq:6.2}
\end{equation}
A small ordering post-Newtonian parameter is defined to be
\begin{equation}
  \gamma = {Gm\over rc^2}  \label{eq:6.3}
\end{equation}
with $r=|{\bf x}|$. Next we use the fact that the origin of the coordinate
system is located at the center of mass of the binary.  This means that
$M_i=0$ where $M_i$ is the dipole mass moment of the external field.  By
Eqs.~(\ref{eq:4.25}) and (\ref{eq:4.27b}) this means ${\widetilde I}_i +
\Delta I_i= O(6)$ where ${\widetilde I}_i$ is the dipole moment which was
computed in paper~II ({\it before} expressing it in the relative frame)
and where $\Delta I_i =(4G/5c^5)mG^{(2)}_i +O(7)$ where $G_i=\int d^3{\bf x}
\,(\sigma_i {\bf x}^2 -{1\over 2} \sigma_j x_ix_j)$.  For circular
orbits one finds $G_i =m\nu (X_2-X_1)r^2v^i + O(2)$ and thus $\Delta
I_i= (4G/5c^5) Gm^3\nu (X_1-X_2)(v^i/r)+O(7)$. This readily shows how the
relations (3.7) of paper~II are to be extended to 2.5PN order. We find
\begin{subequations}
\label{eq:6.4}
\begin{eqnarray}
 y^i_1 &=& [X_2 +3\nu \gamma^2 (X_1-X_2)] x^i - {4\over 5} {G^2m^2\nu\over
  rc^5} (X_1-X_2) v^i + O(6)\ , \label{eq:6.4a}\\
 y^i_2 &=& [-X_1 +3\nu \gamma^2 (X_1-X_2)] x^i - {4\over 5} {G^2m^2\nu\over
  rc^5} (X_1-X_2) v^i + O(6)\ . \label{eq:6.4b}
\end{eqnarray}
\end{subequations}
This result is in agreement with the 2.5PN-accurate center of mass theorem
of Refs.~\cite{DD}.
The assumption that the orbit is circular apart from the adiabatic
inspiral due to reaction effects of order $O(5)$ implies that the scalar
product of $x^i$ and $v^i$ is of small $O(5)$ order:  ${\bf x}.{\bf v}
\equiv (xv)=O(5)$.  By Eqs.~(\ref{eq:6.4}) this implies also
$(nv_1)=O(5)$ and $(nv_2)=O(5)$ where $n^i \equiv x^i/r$.  These facts
simplify drastically the equations of motion of the binary given by
Eqs.~(B1)-(B2) in Appendix~B.  The result when expressed in the
relative frame simply reads
\begin{equation}
 {dv^i\over dt} = -\omega^2_{\rm 2PN} x^i-{32\over 5c^5}{G^3m^3\nu\over r^4}
  v^i + O(6)\ , \label{eq:6.5}
\end{equation}
where we have introduced the angular frequency $\omega_{\rm 2PN}$ defined by
\begin{equation}
 \omega^2_{\rm 2PN} \equiv {Gm\over r^3} \left[ 1-(3-\nu) \gamma +
 \left( 6 +{41\over 4} \nu +\nu^2 \right) \gamma^2 \right]\ . \label{eq:6.6}
\end{equation}
This frequency represents the orbital frequency of the {\it exact} circular
periodic orbit at the 2PN order (see Eq.~(3.11) in paper~II). The relation
between the norm of the relative velocity $v=|{\bf v}|$ and $\omega_{\rm
2PN}$ is obtained by multiplying both sides of (\ref{eq:6.5}) by $x^i$.
Using $(xv)=O(5)$ and $d(xv)/dt=O(10)$ (because $d(xv)/dt$ is of the same
order as the square of reaction effects) we find
\begin{equation}
 v = r\,\omega_{\rm 2PN} + O(6)\ . \label{eq:6.7}
\end{equation}
Finally we write the result for the orbital energy $E\equiv E^{\rm 2.5PN}$
entering the left side of the energy balance equation $dE/dt =-{\cal L}^N
+O(6)$ derived in Eq.~(\ref{eq:B11}) of Appendix~B.  This energy is
computed from Eqs.~(\ref{eq:B7}) and (\ref{eq:B12}) in which one uses
the circular orbit assumption together with (\ref{eq:6.7}). The result is
\begin{equation}
 E = -{c^2\over 2} m\nu \gamma \left\{ 1-{1\over 4} (7-\nu) \gamma
  - {1\over 8} (7-49\nu -\nu^2) \gamma^2 \right\} + O(6)\ . \label{eq:6.8}
\end{equation}
There are no terms of order $\gamma^{5/2} = O(5)$ for circular orbits
because the term $O(5)$ in Eq.~(B11) is proportional to $(nv)$ and thus
vanishes in this case.

\subsection{The energy loss rate} \label{sec:6.2}

The gravitational luminosity ${\cal L}$ of a general source was split into
two contributions, an instantaneous one ${\cal L}_{\rm inst}$ given by
Eq.~(\ref{eq:5.11}) and a tail one ${\cal L}_{\rm tail}$ given by
Eq.~(\ref{eq:5.12}). We shall basically show that only ${\cal L}_{\rm
tail}$ contributes to the 2.5PN order in the case (but only in this case) 
of a binary system moving on a {\it circular} orbit.

Let us consider first the contribution ${\cal L}_{\rm inst}$.
The only moment it contains which is required with full 2.5PN accuracy
is the mass quadrupole moment
\begin{equation}
 M_{ij} = {\widetilde I}_{ij} + \Delta I_{ij} + O(6) \label{eq:6.9}
\end{equation}
where ${\widetilde I}_{ij}$ results from Eq.~(\ref{eq:4.8a}) and $\Delta
I_{ij}$ is given by Eq.~(\ref{eq:4.27c}).  For a circular orbit,
the moment of inertia $K$ is constant [indeed $K=m\nu r^2 +O(2)$], hence
$\Delta I_{ij}$ reduces to two terms only:
\begin{equation}
 \Delta I_{ij} = {G\over c^5} \left\{ - {2\over 3} K Q^{(3)}_{ij}
  - {8\over 7} Q^{(3)}_{k<i} Q_{j>k} \right\} + O(7)\ . \label{eq:6.10}
\end{equation}
We prove that the contribution in ${\cal L}_{\rm inst}$ which is due to
$\Delta I_{ij}$ is in fact zero.  Indeed this contribution is made of
the contracted product between $Q^{(3)}_{ij}$ and $\Delta I^{(3)}_{ij}$
(recall that $M_{ij}=Q_{ij}+O(2)$), hence of the contracted products
$Q^{(3)}_{ij} Q^{(6)}_{ij}$ and $Q^{(3)}_{ij} Q^{(6-q)}_{ik} Q^{(q)}_{jk}$.
But in the circular case an odd number of time-derivatives of $Q_{ij}$
yields a term proportional to $x^{<i}v^{j>}$ while an even number yields
either $x^{<i} x^{j>}$ or $v^{<i}v^{j>}$. Thus a contracted product of the
type $Q^{(n)}_{ij} Q^{(p)}_{ij}$ where $n+p$ is {\it odd} necessarily
involves one scalar product $(xv)$ and is thus zero; similarly a product of
the type $Q^{(n)}_{ij} Q^{(p)}_{ik} Q^{(q)}_{jk}$ where $n+p+q$ is odd is
also zero. The same is true of a product like $\varepsilon_{ijk} Q^{(r)}_{jm}
Q^{(s)}_{km}$ where $r+s$ is {\it even}. These simple facts show that
(\ref{eq:6.10}) cannot contribute to the energy loss rate. Furthermore we
find that all the terms in ${\cal L}_{\rm inst}$ which involve the contracted
products of {\it three} moments [i.e., all the terms of order $c^{-5}$ in
Eq.~(\ref{eq:5.11})]  are also zero.  Hence we can write in the circular
orbit case
\begin{eqnarray}
 {\cal L}_{\rm inst} &=& {G\over c^5} \left\{ {1\over 5}
  {\widetilde I}^{(3)}_{ij} {\widetilde I}^{(3)}_{ij} + {1\over c^2}
  \left[ {1\over 189} {\widetilde I}^{(4)}_{ijk}
 {\widetilde I}^{(4)}_{ijk}+ {16\over 45} {\widetilde J}^{(3)}_{ij}
  {\widetilde J} ^{(3)}_{ij} \right] \right.  \nonumber\\
   && \qquad\quad \qquad \left. + {1\over c^4} \left[ {1\over 9072}
  {\widetilde I}^{(5)}_{ijkm} {\widetilde I}^{(5)}_{ijkm}
 + {1\over 84} {\widetilde J}^{(4)}_{ijk} {\widetilde J}^{(4)}_{ijk}
      \right] + O(6) \right\}   \ .\label{eq:6.11}
\end{eqnarray}
Now recall that the moments ${\widetilde I}_L$ and ${\widetilde J}_L$ are
the ones which were used as starting point in the computation of paper~II.
In using paper~II one must be careful that in this paper the equations
of motion have a precision limited to 2PN instead of 2.5PN. We first note
that the 2.5PN terms in the mass-centered-frame relations (\ref{eq:6.4})
will cancel out when expressing the quadrupole mass moment in terms of
the relative variables. Furthermore during the computation of this moment
in paper~II the equations of motion were used only to reduce some terms
which were already of 1PN order. Thus all the expressions of the moments
$\widetilde I_L$ and $\widetilde J_L$ computed in paper~II (but not the
expressions of their time derivatives) can be used in the present paper
without modification.
Notably the mass quadrupole ${\widetilde I}_{ij}$ takes the expression
(3.74) of paper~II, modulo negligible $O(6)$ terms:
\begin{eqnarray}
 {\widetilde I}_{ij} &=& {\rm STF}_{ij}\, \nu m \biggl\{ x^{ij}
  \left[ 1 - {\gamma\over 42}
  (1+39\nu)-{\gamma^2\over 1512} (461+18395\nu +241\nu^2) \right]\nonumber\\
  &&\qquad\qquad\quad +{r^2\over c^2} v^{ij} \left[ {11\over 21}
  (1-3\nu)+{\gamma\over 378} (1607-1681\nu +229\nu^2) \right]\biggr\}
  + O(6)\ . \label{eq:6.12}
\end{eqnarray}
However we need the third time derivative of (\ref{eq:6.12}) which does
involve, because of the more precise equations of motion, some new terms
with respect to paper~II.  We find
\begin{eqnarray}
 {\widetilde I}^{(3)}_{ij} &=&{\rm STF}_{ij}\,\nu m \biggl\{- 8{Gm \over r^3}  x^{i} v^{j} \left[ 1
 -{\gamma\over 42}(149-69\nu)+{\gamma^2\over 1512}(7043-7837\nu +3703\nu^2)
  \right]  \nonumber\\
 && \qquad\qquad\quad +{64\over 5}{G^3m^3\nu\over r^4c^5}\left[ {Gm\over r^3}
     x^{ij} - 3 v^{ij} \right] \biggr\} + O(6)\ .  \label{eq:6.13}
\end{eqnarray}
But for the same reason as before these new terms will not contribute to
the energy loss for circular orbits.  Thus we conclude that ${\cal
L}_{\rm inst}$ is exactly given by the expression found in Eq.~(4.12) of
paper~II modulo negligible terms of relative order $O(6)\equiv
O(\gamma^3)$.  We can thus write
\begin{equation}
  {\cal L}_{\rm inst} = {32c^5\over 5G} \nu^2 \gamma^5 \biggl\{
  1 - \left( {2927\over 336} + {5\over 4} \nu \right) \gamma
  + \left( {293383\over 9072} + {380\over 9} \nu \right) \gamma^2
  + O(\gamma^3) \biggr\}\ . \label{eq:6.14}
\end{equation}

Let us now turn our attention to the tail part of the gravitational
luminosity as defined by Eq.~(\ref{eq:5.12}). The correct formula to
compute the tail integrals in ${\cal L}_{\rm tail}$ is
\begin{equation}
 \int^{+\infty}_0 d\tau\ln \left( {\tau\over 2b} \right) \cos (\Omega \tau)
  = - {\pi\over 2\Omega} \ , \label{eq:6.15}
\end{equation}
where $\Omega$ denotes the (real) angular frequency of the radiation. It
was proved in Ref.~\cite{BS93} that this formula is to be applied as it
stands (i.e., even though the integral is not absolutely convergent) to a
fixed (non-decaying) orbit whose frequency is equal to the current value
of the orbital frequency $\omega \equiv \omega_{\rm 2PN} (T_R)$. The
numerical errors done in applying this formula were shown to be of relative
order $O(\ln c /c^5)$, which is always negligible for our purpose here
(because of the explicit powers of $c^{-1}$ in front of the tail
integrals).  Thus we replace into the integrand of (\ref{eq:5.12}) the
moments $M_L$ and $S_L$ by their expressions valid for fixed circular
orbits and up to the appropriate precision.  The necessary formulas
(taken from paper~II) are
\begin{subequations}
 \label{eq:6.16}
\begin{eqnarray}
 M^{(3)}_{ij} &=& -8\,{\rm STF}_{ij}\,{Gm^2\nu\over r^3} x^{i}v^{j} \left[
  1-{\gamma\over 42} (149-69\nu) \right] +O(4)\ , \label{eq:6.16a}\\
 M^{(5)}_{ij} &=& 32\, {\rm STF}_{ij}\,{G^2m^3\nu\over r^6} x^{i}v^{j} \left[
  1-{\gamma\over 42} (275-111\nu) \right] +O(4)\ , \label{eq:6.16b}\\
 M^{(4)}_{ijk} &=& 3 (X_2-X_1)\,{\rm STF}_{ijk}\,{Gm^2\nu\over r^3}
  \left\{ 7 {Gm\over r^3} x^{ijk} - 20 x^{i}v^{jk} \right\}
    +O(2)\ , \label{eq:6.16c}\\
 M^{(6)}_{ijk} &=& 3 (X_2-X_1)\,{\rm STF}_{ijk}\,{G^2m^6\nu\over r^6}
  \left\{ -61 {Gm\over r^3} x^{ijk} +  182 x^{i}v^{jk} \right\}
    +O(2)\ , \label{eq:6.16d}\\
 S^{(3)}_{ij} &=& - (X_2-X_1)\,{\rm STF}_{ij}\,{Gm^2\nu\over r^3}
  \varepsilon^{abi} v^{j} x^a v^b + O(2)\ , \label{eq:6.16e}\\
 S^{(5)}_{ij} &=& (X_2-X_1)\,{\rm STF}_{ij}\,{G^2m^3\nu\over r^6}
  \varepsilon^{abi} v^{j}  x^a v^b + O(2)\ . \label{eq:6.16f}
\end{eqnarray}
\end{subequations}
The quadrupole moment includes 1PN terms while the other moments are
Newtonian.  The computation of ${\cal L}_{\rm tail}$ involves many scalar
products between $x^i$ or $v^i$ given at the current time $T_R$,
i.e.  ${x^i}\equiv {x^i}(T_R)$ or ${v^i}\equiv {v^i}(T_R)$, with the
same vectors but evaluated at an arbitrary earlier time $T_R-\tau$, say
${{x'}^{i}}\equiv {x^{i}}(T_ R-\tau)$ and ${{v'}^{i}}\equiv {v^{i}}
(T_R-\tau)$.  Relevant formulas for these scalar products are
\begin{subequations}
\label{eq:6.17}
\begin{eqnarray}
 (xx') &=& r^2 \cos (\omega\tau)\ , \label{eq:6.17a}\\
 (vv') &=& r^2 \omega^2 \cos (\omega\tau)\ , \label{eq:6.17b}\\
 (xv') &=& -(vx') = r^2 \omega \sin (\omega\tau)\ , \label{eq:6.17c}
\end{eqnarray}
\end{subequations}
where $\omega$ is the current value of the frequency. The reduction of
${\cal L}_{\rm tail}$ is quite straightforward. We need to use
$\omega^2 =(Gm/r^3)[1-(3-\nu)\gamma +O(\gamma^2)]$ (the 1PN correction
in $\omega^2$ is sufficient) and the formula (\ref{eq:6.15}) where
$\Omega = \omega$, $2\omega$ or $3\omega$. One uses also $(X_2-X_1)^2
=1-4\nu$. The result reads
\begin{equation}
  {\cal L}_{\rm tail} = {32c^5\over 5G}  \nu^2 \gamma^5 \left\{ 4\pi
      \gamma^{3/2} - \left( {25663\over 672} + {125\over 8} \nu \right)
      \pi \gamma^{5/2} + O(\gamma^{3}) \right\} \ . \label{eq:6.18}
\end{equation}

The complete 2.5PN-accurate gravitational luminosity generated by a
inspiralling compact binary moving on a quasi-circular orbit is therefore
obtained as the sum of Eqs.~(\ref{eq:6.14}) and (\ref{eq:6.18}). We obtain
\begin{eqnarray}
 {\cal L} &=& {32c^5\over 5G} \nu^2 \gamma^5 \biggl\{
   1 - \left( {2927\over 336} + {5\over 4} \nu\right) \gamma +
   4 \pi\,\gamma^{3/2} + \left( {293383\over 9072} +{380\over 9}\nu \right)
     \gamma^2 \nonumber \\
   &&\qquad\qquad\qquad - \left( {25663\over 672} + {125\over 8}\nu \right)
     \pi \gamma^{5/2} + O(\gamma^3) \biggr\}\ , \label{eq:6.19}
\end{eqnarray}
where we recall that the post-Newtonian ordering parameter is $\gamma=Gm/rc^2$
with $r$ being the harmonic-coordinate orbital separation. This expression
was already obtained to 1PN order in Refs.~\cite{EW75,WagW76,BS89}, to
1.5PN order in Refs.~\cite{BD92,P93,Wi93,BS93}, and to 2PN order in
Refs.~\cite{B95,BDI95,WWi95,BDIWWi95}. The 2.5PN order added in this paper
is like the 1.5PN order due to the presence of the radiation tails in the
wave zone, as indicated by the irrational number $\pi$ in factor coming from
the formula (6.15).  Hence there is up to 2.5PN order a clean separation
between the integer post-Newtonian approximations which come from
instantaneous relativistic effects in the source multipole moments and
the half-integer approximations which are due to hereditary effects in
the wave zone.  However the next post-Newtonian approximation (3PN) is
expected to involve both types of effects.

\subsection{The orbital phase} \label{sec:6.3}

Once the energy loss has been derived in a particular coordinate system
it can be re-expressed in a coordinate-independent way by using the
directly observable frequency $\omega \equiv \omega_{\rm 2PN}$ instead
of the parameter $\gamma$.  Defining $x=(Gm\omega /c^3)^{2/3}$, we
obtain the inverse of Eq.~(\ref{eq:6.6}) as
\begin{equation}
 \gamma = x \left[ 1+ \biggl( 1-{\nu \over 3}\biggr) x
  +\left( 1-{65\nu\over 12} \right) x^2 + O(x^3) \right]\ ,\label{eq:6.20}
\end{equation}
which is substituted into Eq.~(\ref{eq:6.19}) with the result
\begin{eqnarray}
 {\cal L} &=& {32c^5\over 5G} \nu^2 x^5 \biggl\{ 1
  - \left( {1247\over 336}+{35\over 12}\nu \right) x + 4\pi x^{3/2}
  + \left(-{44711\over 9072}+{9271\over 504}\nu + {65\over 18} \nu^2
    \right) x^2 \nonumber \\
 &&\quad\qquad\qquad -\left({8191\over 672}+{583\over 24}\nu\right)\pi x^{5/2}
    + O(x^3) \biggr\}  \ . \label{eq:6.21}
\end{eqnarray}
An important check is obtained in the test-body limit $\nu\to 0$ of this
result which is found to agree with the 2.5PN truncation of the result of
perturbation theory known through the equivalent of 4PN order
\cite{TNaka94,Sasa94,TSasa94} (see the equation (43) of Ref.\cite{TSasa94}).
As for the orbital energy $E\equiv E^{\rm 2.5PN}$ which has been given
in Eq.~(\ref{eq:6.8}) following the equations of motion summarized in
Appendix~B, it reads in terms of $x$ as
\begin{equation}
 E = -{c^2\over 2} m\nu x \left\{ 1 - {1\over 12} (9+\nu) x -{1\over 8}
 \left(27-19\nu +{\nu^2\over 3}\right) x^2 + O(x^3) \right\}\ .\label{eq:6.22}
\end{equation}

The laws of variation of the frequency $\omega$ and phase $\phi$ of the
binary during the orbital decay will be computed using the energy
balance equation
\begin{equation}
 {dE\over dT_R} = - {\cal L}\ , \label{eq:6.23}
\end{equation}
where $T_R$ is the retarded time in the wave zone.
Note that we meet here the remarks made in the introduction. Up to now
the analysis has been rigorous within the framework of post-Newtonian
theory. Namely, ${\cal L}$ has been computed from a well-defined formalism
based on convergent integrals for the multipole moments (see Section
IV). The reduction of ${\cal L}$ to binary systems using delta functions
to describe the compact objects
can probably be justified at the 2.5PN order by using the results of
Ref.~\cite{D83a} (see paper II). Furthermore $E$ is computed directly from
the Damour-Deruelle equations of motion \cite{DD}. We now {\it postulate} the
validity of the energy balance equation (\ref{eq:6.23}) where both $E$ and
${\cal L}$ take their full 2.5PN accuracy. This equation has been proved
here (in Eq.~(\ref{eq:B10}) of Appendix B) only when ${\cal L}$ takes its
Newtonian value ${\cal L}^N$.  It was proved before to Newtonian order
in Refs.~\cite{CE70,Ker80,PaL81,BRu81,A87,DD}.  On the other hand the
balance equation is also known to hold at 1PN order \cite{IW,B'}, and even
for tails at 1.5PN order \cite{BD88,BD92}. To prove it at 2.5PN order like
in (\ref{eq:6.23}) would mean knowing all the radiation reaction effects
in the binary's equations of motion up to the 5PN order.  To palliate
this it has been customary in this field (see
Refs.~\cite{3mn,FCh93,CF94,P93,CFPS93,P95,CF95,BDI95}) to compute
the binary's orbital decay assuming that the energy balance equation is
valid to high order (the angular momentum balance equation is
unnecessary for circular orbits).

Adopting the same
approach to the problem, we introduce like in paper~II the adimensional time
\begin{equation}
 \theta = {c^3\nu\over 5Gm} T_R\ . \label{eq:6.24}
\end{equation}
Equation~(\ref{eq:6.23}) is then transformed into the ordinary
differential equation
\begin{eqnarray}
 {dx\over d\theta} &=& 64 x^5 \biggl\{ 1 - \left( {743\over 336} +
  {11\over 4}\nu \right) x + 4\pi x^{3/2} \nonumber \\
  && \qquad\qquad + \left( {34103\over 18144} + {13661\over 2016} \nu
   + {59\over 18} \nu^2 \right) x^2 \nonumber\\
  && \qquad\qquad -\left( {4159\over 672} +{189\over 8} \nu\right)
     \pi x^{5/2} + O(x^3)\biggr\}\ . \label{eq:6.25}
\end{eqnarray}
On the other hand the differential equation for the instantaneous phase
$\phi$ of the binary is $d\phi =\omega dT_R = (5/\nu) x^{3/2} d\theta$ or
equivalently
\begin{eqnarray}
 {d\phi\over dx} &=& {5\over 64\nu} x^{-7/2} \biggl\{ 1 +\left({743\over 336}
 + {11\over 4}\nu \right) x - 4\pi x^{3/2} \nonumber \\
  && \qquad\qquad + \left( {3058673\over 1016064} + {5429\over 1008} \nu
   + {617\over 144} \nu^2 \right) x^2 \nonumber\\
  && \qquad\qquad -\left( {7729\over 672} -{13\over 8} \nu\right)
     \pi x^{5/2} + O(x^3)\biggr\}\ . \label{eq:6.26}
\end{eqnarray}
The frequency and phase follow from the integration of
these equations.  The frequency (or $x$-parameter) reads
\begin{eqnarray}
 x &=& {1\over 4} \Theta^{-1/4} \biggl\{ 1 + \left( {743\over 4032}
 +{11\over 48}\nu\right)\Theta^{-1/4} - {\pi\over 5}\Theta^{-3/8}\nonumber\\
  && \qquad\qquad + \left( {19583\over 254016} + {24401\over 193536} \nu
   + {31\over 288} \nu^2 \right) \Theta^{-1/2} \nonumber\\
  && \qquad\qquad +\left(-{11891\over 53760} +{109\over 1920}\nu\right)
     \pi \Theta^{-5/8} + O(\Theta^{-3/4} )\biggr\}\ , \label{eq:6.27}
\end{eqnarray}
where $\Theta = \theta_c - \theta$ denotes the (adimensional) time
left till the final coalescence, and the phase is
\begin{eqnarray}
 \phi &=& -{1\over 32\nu} \biggl\{ x^{-5/2} +\left({3715\over 1008}
 + {55\over 12}\nu \right) x^{-3/2} - 10\pi x^{-1} \nonumber \\
  && \qquad\qquad + \left( {15293365\over 1016064} + {27145\over 1008} \nu
   + {3085\over 144} \nu^2 \right) x^{-1/2} \nonumber\\
  && \qquad\qquad +\left( {38645\over 1344} -{65\over 16} \nu\right)
     \pi \ln \left({x\over x_0}\right) + O(x^{1/2})\biggr\}\ ,\label{eq:6.28}
\end{eqnarray}
where the constant $x_0$ is determined by initial conditions.
By substituting Eq.~(\ref{eq:6.27}) into (\ref{eq:6.28}) one obtains the
phase as a function of time:
\begin{eqnarray}
\phi &=& - {1\over \nu} \biggl\{ \Theta^{5/8} + \left(
  {3715\over 8064} + {55\over 96} \nu \right) \Theta^{3/8}
    - {3\pi\over 4} \Theta^{1/4} \nonumber\\
  &&\qquad\qquad + \left( {9275495\over 14450688} + {284875\over 258048} \nu
    + {1855\over 2048} \nu^2  \right) \Theta^{1/8} \nonumber \\
  &&\qquad\qquad - \left( {38645\over 172032} - {65\over 2048} \nu \right)
  \pi \ln \left({\Theta\over \Theta_0}\right) + O (\Theta^{-1/8}) \biggr\}
   \ ,  \label{eq:6.28'}
\end{eqnarray}
where $\Theta_0$ denotes some constant.

The presence of the logarithm of the frequency or coalescing time
in the phase is a novel
feature of the 2.5PN approximation.  As a result the phase $\phi$ tends
formally to infinity at the time of coalescence.  In other words, the
number of cycles left till the time of coalescence starting from an
initial frequency is always infinite. This is in contrast with the previous
2PN approximation where $\phi$ was tending to a finite value $\phi_c$.
Of course this peculiarity is not relevant to the actual physics of the
binary because the post-Newtonian approximation breaks down before the
coalescence, at some finite value of the frequency.  The fact
that $\phi$ tends to infinity is merely a consequence of our idealized
model assuming that the two bodies have no internal structure or horizons,
which is valid only during the inspiral phase preceding
the coalescence. However, during this phase the model is expected to be
very precise and in particular the logarithmic 2.5PN term in the phase
has to be taken numerically into account.  This logarithm makes in fact
the contribution of the 2.5PN term to the total number of cycles in the
detector's bandwidth somewhat more important than what could be deduced
from a simple order of magnitude estimate.

Let us be more precise and evaluate the contribution of each post-Newtonian
terms in the phase (\ref{eq:6.28}) to the accumulated number ${\cal N}$ of
gravitational-wave cycles between some initial and final frequencies $\omega_i$
and $\omega_f$. Note that such a computation can only be indicative of the
relative orders of magnitude of the different terms in the phase. A full
analysis would require the knowledge of the power spectral density of
the noise in a detector, and a complete simulation of the parameter
estimation using matched filtering \cite{3mn,FCh93,CF94,PW95,KKS95}. The
contribution to ${\cal N}$ due to the 2.5PN term in (\ref{eq:6.28}) is
\begin{equation}
 {\cal N}_{\rm 2.5PN} = - {1\over 48\nu} \left( {38645\over 1344}
   - {65\over 16}\nu \right) \ln \left( {\omega_f\over \omega_i}\right)
   \ . \label{eq:6.29}
\end{equation}
In the case of the inspiral of two neutron stars of mass $1.4M_\odot$,
and with the values $\omega_i/\pi =10$Hz (set by seismic noise) and $\omega_f
/\pi =1000$Hz (set by photon shot noise), we find from Ref.~\cite{BDIWWi95}
and Eq.~(\ref{eq:6.29})

\[ \begin{tabular}{c|c|c|c|c|c}
  &\quad Newtonian      &\quad 1PN &\quad 1.5PN  &\quad 2PN
  &\quad 2.5PN \\ & & & & &\\
\hline & & & & & \\
  ${\cal N}$ &16,050 &439  &--208  &9  &--11 \\
\end{tabular}  \]
[Note that as far as the final frequency $\omega_f$ is determined only by
detector's characteristics (and not by the location of the
innermost stable orbit),
the term ${\cal N}_{\rm 2.5PN}$ depends on the masses only through the
mass ratio $\nu$.]

In this (indicative) example we see that the contribution of the 2.5PN order
more than compensates the contribution of the previous 2PN order. [But of
course both contributions have to be included in the filters since they have
different functional dependences on the frequency.] However note that
finite mass effects (proportional to $\nu$) are numerically small
at the 2.5PN order, contrarily to the 2PN order where they are quite
significant \cite{BDI95,BDIWWi95}. In the previous example, they contribute
only to $-0.1$ cycle as compared to ${\cal N}_{\rm 2.5PN} =-11$.

\appendix
\section{The proof of some formula}\label{sec:apa}

To prove the formula (\ref{eq:4.19}) we begin with the identity
\begin{equation}
 r^B \Box f = \Box (r^Bf) - B(B+1) r^{B-2} f - 2Br^{B-1} \partial_r
   f \label{eq:A1}
\end{equation}
which permits the transformation of the left side of (\ref{eq:4.19}), namely
\begin{equation}
 {}_PX_Q \equiv ({\rm FP}_{B=0} \Box^{-1}_R r^B \Box -\openone) \left\{
  \hat\partial_P \left[ {1\over r} F (t-r/c)\right] \hat\partial_Q
  \left[ {1\over r} G (t-r/c)\right] \right\} \ , \label{eq:A2}
\end{equation}
into
\begin{equation}
 {}_PX_Q = {\rm FP}_{B=0} \Box^{-1}_R \left( -Br^{B-1} [2\partial_r +r^{-1}]
 \left\{ \hat\partial_P \left[ {1\over r} F (t-r/c)\right] \hat\partial_Q
  \left[ {1\over r} G (t-r/c)\right] \right\} \right)\ . \label{eq:A3}
\end{equation}
Because of the presence of the explicit factor $B$ in the integrand, the
equation (\ref{eq:A3}) appears to be the {\it residue} of the Laurent expansion
of the retarded integral near $B=0$ (we have discarded a term proportional
to $B^2$ which is known to give zero contribution). Residues of retarded
integrals have been investigated in Eq.~(4.26) of Ref.~\cite{BD88} and
yield retarded solutions of the wave equation. By a simple dimensional
analysis we find that ${}_PX_Q$ admits the structure
\begin{equation}
 {}_PX_Q \sim \sum^{[{(p+q)/2}]}_{k=0} {1\over c^{2k+1}} \hat\partial_L
 \left[ {1\over r} F^{(a)} (t-r/c) G^{(b)} (t-r/c) \right]\ . \label{eq:A4}
\end{equation}
Here, $[(p+q)/2]$ is the integer part of $(p+q)/2$,
$\hat\partial_L$ denotes a tracefree derivative operator composed of $\ell$
spatial derivatives where $\ell=p+q-2k$, and the number of time derivatives
on $F$ and $G$ is $a+b=2k+1.$ Omitted in (A4) are pure dimensionless
coefficients and numerous Kronecker symbols. The explicit dependence on $1/c$
present in Eq.~(\ref{eq:A4}) shows that if we are interested only in the
leading-order term when $c\to +\infty$ we can limit us to
the computation of the terms having the {\it maximum} number of space
derivatives, i.e.  $\ell_{\max} =p+q$ (corresponding to $k = 0$). The errors
made in keeping only such terms are of relative order $O(3)$.  We
use to compute (\ref{eq:A3}) the formula
\begin{equation}
\hat\partial_P \left[ {1\over r} F(t-r/c)\right] = \hat n_P
\sum^p_{i=0} a^p_i {F^{(p-i)}(t-r/c)\over c^{p-i} r^{1+i}}\ , \label{eq:A5a}
\end{equation}
where the numerical coefficient is
\begin{equation}  
a^p_i = {(-)^p (p+i) !\over 2^i i! (p-i)!} \  \label{eq:A5b}
\end{equation}
(see Eq.~(A35a) in \cite{BD86}). This yields a double sum of terms having
in front the product of the tracefree tensors $\hat n_P$ and $\hat n_Q$. This
product can be decomposed into tracefree tensors of multipolarities
$\ell =p+q-2k$ with $k = 0, ..., [(p+q)/2]$, however if we are looking
only for the term having $\ell_{\max}=p+q$ one can simply replace the
product $\hat n_P \hat n_Q$ by the tracefree tensor $\hat
n_{PQ}$ (with one as a coefficient in front).  Hence
\begin{equation}
 {}_PX_Q = \sum_{i,j} a^p_i a^q_j {\rm FP}_{B=0}
    \Box^{-1}_R \left\{ -Br^{B-1} \hat n_{PQ} [2\partial_r +r^{-1}]
  {F^{(p-i)}(t-r/c) G^{(q-j)}(t-r/c)\over c^{p+q-i-j} r^{2+i+j}} \right\}
  + O(3)\ ,  \label{eq:A6}
\end{equation}
where the remainder $O(3)$ comes from the errors made in the replacement
$\hat n_P\hat n_Q\to \hat n_{PQ}$. The integrand in (\ref{eq:A6}) is easily
transformed into a sum of terms of the form $r^{B-K} \hat n_{PQ} H(t-r/c)$
times the factor $B$, where $K$ is an integer. Using the fact that the
residues of such terms exist only when $K\geq p+q+3$ (see Eq.~(4.26) in
\cite{BD88}), one finds that the summation indices $i$ and $j$ in (\ref{eq:A6})
must satisfy $i+j =p+q$ or $p+q-1$. This simplifies much the computation
of (\ref{eq:A6}) which is done using the expression  (A6) of the
coefficients $a^p_i$ and the equation (4.26) in \cite{BD88}. The result is
\begin{equation}
 {}_PX_Q = \hat\partial_{PQ}
 \left[ {1\over rc} \biggl(\delta_{p,0} F^{(1)} G
  + \delta_{0,q} FG^{(1)} \biggr) \right] + O(3)\ , \label{eq:A7}
\end{equation}
where $\delta_{p,0}$ denotes the usual Kronecker symbol.

\section{The Damour--Deruelle equations of motion}  \label{sec:apb}

We present a summary of results concerning the equations of motion of two
point-masses moving under their mutual gravitational influence up to the
2.5PN order.  These equations were obtained by Damour and Deruelle
\cite{DD} (see also the presentation
\cite{D83a} of which we are close here).  The equations
of motion of body 1 (say) take the Newtonian-like form
\begin{subequations}
\label{eq:B1}
\begin{eqnarray}
 {dy^i_1 \over dt} &=& v^i_1\ , \label{eq:B1a}   \\
 {dv^i_1 \over dt} &=& A^i_1 + {1\over c^2} B^i_1 + {1\over c^4} C^i_1
   + {1\over c^5} D^i_1 + O(6)\ , \label{eq:B1b}
\end{eqnarray}
\end{subequations}
where $y^i_1$ and $v^i_1$ denote the instantaneous position and coordinate
velocity of body 1 (in the harmonic coordinate system), $A^i_1$ is the
usual Newtonian acceleration of the body, and $B^i_1$, $C^i_1$ and $D^i_1$
represent the relativistic corrections of order 1PN, 2PN and 2.5PN
respectively. The equations (12.7)-(12.12) in Ref.~\cite{D83a} give
these terms as
\begin{subequations}
\label{eq:B2}
\begin{eqnarray}
 A^i_1 &=& - {Gm_2\over r^2} n^i\ , \label{eq:B2a}\\
 B^i_1 &=& {Gm_2\over r^2} \biggl\{ n^i \left[ -v^2_1 - 2v^2_2 + 4(v_1v_2)
  + {3\over 2} (nv_2)^2 + 5{Gm_1\over r} + 4{Gm_2\over r} \right]\nonumber\\
 && \qquad\qquad +(v^i_1-v^i_2) \biggl[4(nv_1) -3(nv_2)\biggr] \biggr\}\ ,
      \label{eq:B2b}\\
 C^i_1 &=& {Gm_2\over r^2} \biggl\{ n^i \biggl[ -2v^4_2 + 4v^2_2 (v_1v_2)
  - 2(v_1v_2)^2  + {3\over 2} v^2_1 (nv_2)^2 \nonumber\\
 &&\qquad\qquad +{9\over 2} v^2_2 (nv_2)^2 - 6(v_1v_2) (nv_2)^2
  - {15\over 8} (nv_2)^4     \nonumber\\
 && \qquad\quad+ {Gm_1\over r} \biggl( -{15\over 4} v^2_1 +{5\over 4} v^2_2
    -{5\over 2} (v_1v_2) + {39\over 2} (nv_1)^2
    -39(nv_1)(nv_2)+{17\over 2}(nv_2)^2 \biggr)\nonumber\\
 && \qquad\quad +{Gm_2\over r}\biggl( 4 v^2_2 - 8(v_1v_2)+ 2(nv_1)^2
  - 4(nv_1)(nv_2) - 6(nv_2)^2\biggr) \biggr] \nonumber \\
 && \qquad\quad +(v_1^i -v^i_2)\biggl[ v^2_1(nv_2)+4v^2_2(nv_1) -5v^2_2(nv_2)
    -4(v_1v_2)(nv_1)\nonumber\\
  && \quad\qquad\qquad + 4(v_1v_2)(nv_2) -6(nv_1)(nv_2)^2
    + {9\over 2} (nv_2)^3\nonumber\\
 && \qquad\qquad +{Gm_1\over r} \left( -{63\over 4}(nv_1) + {55\over 4}
  (nv_2)\right) +{Gm_2\over r} \biggl(-2(nv_1)-2(nv_2)\biggr)\biggr]
    \biggr\} \nonumber\\
 && + {G^3m_2\over r^4} n^i \left\{ -{57\over 4}m^2_1 - 9m^2_2
   - {69\over 2} m_1m_2 \right\}\ ,  \label{eq:B2c} \\
 D^i_1 &=& {4\over 5} {G^2m_1m_2\over r^3} \biggl\{ v^i \left[ -v^2
  + 2{Gm_1\over r} - 8{Gm_2\over r} \right]
  + n^i (nv) \left[ 3v^2 -6{Gm_1\over r} + {52\over 3}
     {Gm_2\over r}\right] \biggr\} \ , \label{eq:B2d}
\end{eqnarray}
\end{subequations}
where $m_1$ and $m_2$ are the two masses, $v^i_1$ and $v^i_2$ the two
velocities, $r=|{\bf y}_1-{\bf y_2}|$ the separation, and
$n^i=(y^i_1-y^i_2)/r$.  Scalar products are denoted by e.g.  ${\bf
v}_1\cdotp {\bf v}_2= (v_1v_2)$.  In the last equation $v^i=v^i_ 1
-v^i_2$.  The equations of the second body are obtained by exchanging the
labels 1 and 2, being careful to the fact that $n^i$ and $v^i$ change
sign in the exchange. The acceleration term $D^i_1$ is responsible for the
dominant damping or radiation reaction effects in the dynamics of the
binary.

The equations of motion when truncated to 2PN order (neglecting the
damping term $D^i_1$) admit a Lagrangian formulation and the associated
conservations laws \cite{DD}. The 2PN Lagrangian in harmonic coordinates
depends not only on the positions and velocities of the two bodies, but
also on their accelerations $a^i_{1,2} =dv^i_{1,2} /dt$: $L^{\rm 2PN}=
L^{\rm 2PN}(y,v,a)$.  It is given by
\begin{eqnarray}
 L^{\rm 2PN} (y,v,a) &=& \sum \biggl\{ {1\over 2} m_1v^2_1 +{1\over 2}
  {Gm_1m_2\over r} + {1\over 8} m_1v^4_1 \nonumber\\
 &&\qquad +{Gm_1m_2\over r} \left[ {3\over 2} v^2_1 - {7\over 4} (v_1v_2)
    -{1\over 4} (nv_1)(nv_2) - {1\over 2} {Gm_1\over r} \right] \nonumber\\
 &&\qquad +{1\over 16} m_1v_1^6 + {Gm_1m_2\over r} \biggl[ {7\over 8} v^4_1
    +{15\over 16} v^2_1v^2_2 -2v^2_1 (v_1v_2) \nonumber\\
 &&\qquad +{1\over 8} (v_1v_2)^2 -{7\over 8}(nv_1)^2v^2_2 + {3\over 4}(nv_1)
      (nv_2)(v_1v_2) +{3\over 16} (nv_1)^2 (nv_2)^2 \biggr] \nonumber\\
 &&\qquad + {G^2m^2_1m_2\over r^2} \biggl[ {1\over 4} v^2_1 +{7\over 4} v^2_2
   -{7\over 4}(v_1v_2) +{7\over 2} (nv_1)^2 + {1\over 2} (nv_2)^2
   - {7\over 2} (nv_1)(nv_2) \biggr] \nonumber\\
 && \qquad+Gm_1m_2 \left[(na_1)\left({7\over 8}v^2_2-{1\over 8}(nv_2)^2\right)
    - {7\over 4} (v_2a_1)(nv_2) \right] \biggr\} \nonumber\\
 && + {G^3m_1m_2\over r^3} \left[ {1\over 2} m^2_1 + {1\over 2} m^2_2
    +{19\over 4} m_1m_2 \right]\ . \label{eq:B4}
\end{eqnarray}
The summation symbol runs over the two bodies 1 and 2. The equations obtained
by variation of this Lagrangian, and in which the accelerations are replaced
{\it after} variation by their (Newtonian) values, are equivalent to the 2PN
equations of motion (Eqs.~(B1)-(B2) where $D^i_1=0$) modulo negligible
$O(6)$ terms. From this Lagrangian one constructs the 2PN integral of energy
\begin{equation}
 \widetilde E^{\rm 2PN} = \sum\left\{v^i_1 \left({\partial L^{\rm 2PN}\over
 \partial v^i_1} - {d\over dt} {\partial L^{\rm 2PN}\over \partial a^i_1}
 \right) + a^i_1 {\partial L^{\rm 2PN}\over \partial a^i_1} \right\}
       - L^{\rm 2PN} \label{eq:B5}
\end{equation}
and one replaces in it the accelerations by their (Newtonian) values,
resulting in
\begin{equation}
 \widetilde E^{\rm 2PN} (y,v,a) = E^{\rm 2PN} (y,v) + O(6)\ . \label{eq:B6}
\end{equation}
The explicit expression of the energy $E^{\rm 2PN}$ as a function of the
positions and velocities is
\begin{eqnarray}
  E^{\rm 2PN} (y,v) &=& \sum \biggl\{ {1\over 2} m_1v^2_1 - {1\over 2}
   {Gm_1m_2\over r} + {3\over 8c^2} m_1 v^4_1 \nonumber\\
 &&\qquad + {Gm_1m_2\over c^2r} \left[ {3\over 2} v^2_1 - {7\over 4} (v_1v_2)
  -{1\over 4} (nv_1)(nv_2) +{1\over 2} {Gm_1\over r}\right]\nonumber\\
 &&\qquad + {5\over 16c^4} m_1v_1^6 +{Gm_1m_2\over c^4r} \biggl[ {21\over 8}
   v^4_1 +{31\over 16} v^2_1 v^2_2 \nonumber\\
 &&\qquad\qquad -{55\over 8}v^2_1 (v_1v_2) +{17\over 8} (v_1v_2)^2
    -{13\over 8}(nv_1)^2v^2_2 \nonumber\\
 &&\qquad\qquad -{9\over 8} (nv_1)(nv_2) v^2_2
  +{3\over 4}(nv_1)(nv_2)(v_1v_2)\nonumber\\
 &&\qquad\qquad +{13\over 8}(v_1v_2)(nv_2)^2 +{3\over 16}(nv_1)^2(nv_2)^2
    +{3\over 8} (nv_1)(nv_2)^3\biggr]\nonumber\\
 &&\qquad +{G^2m^2_1m_2\over c^4r^2} \left[ -{3\over 2} v^2_1 +{7\over 4}v^2_2
   +{29\over 4}(nv_1)^2 +{1\over 2}(nv_2)^2
  -{13\over 4}(nv_1)(nv_2)\right] \biggr\} \nonumber \\
 &&- {G^3m_1m_2\over c^4r^3} \left[ {1\over 2} m^2_1 +{1\over 2} m^2_2
   +{19\over 4}m_1m_2 \right]\ . \label{eq:B7}
\end{eqnarray}
The fact that $\widetilde E^{\rm 2PN}$ is the integral of energy of the 2PN
equations of motion implies that $E^{\rm 2PN}$ as a function of the positions
and velocities satisfies the identity
\begin{equation}
 \sum \left\{ v^i_1 {\partial E\over \partial y^i_1}^{\rm 2PN} +
  \left[ A^i_1+{1\over c^2} B^i_1 +{1\over c^4} C^i_1\right]
   {\partial E\over \partial v^i_1}^{\rm 2PN} \right\} \equiv O(6)\ ,
  \label{eq:B8}
\end{equation}
from which we find the law of variation of $E^{\rm 2PN}$ under the complete
2.5PN dynamics of the binary, namely
\begin{equation}
 {dE\over dt}^{\rm 2PN} = \sum {1\over c^5} D^i_1
   {\partial E\over \partial v^i_1}^{\rm 2PN} + O(6)\ .\label{eq:B9}
\end{equation}
This equation makes it clear how the damping acceleration term $D^i_1$ drives
the variation of the energy (see Eq.~(15.3) in Ref.~\cite{D83a}).  Since
the right side of (\ref{eq:B9}) is a small post-Newtonian term it can be
evaluated by inserting in place of $E^{\rm 2PN}$ its Newtonian value
given by the first two terms in (\ref{eq:B7}).  Thus
\begin{equation}
  {dE\over dt}^{\rm 2PN} = \sum {1\over c^5} m_1 v^i_1 D^i_1 +O(6)\ .
   \label{eq:B10}
\end{equation}
Using the expression (\ref{eq:B2d}) of the damping term
we arrive after a short calculation at the balance-like equation
\begin{equation}
  {dE\over dt}^{\rm 2.5PN} = - {\cal L}^{\rm N} + O(6)\ , \label{eq:B11}
\end{equation}
which involves in the left side the 2PN energy (B6) augmented by
a pure 2.5PN term, namely
\begin{equation}
 E^{\rm 2.5PN} = E^{\rm 2PN} + {8\over 5c^5} {G^2m^3\nu^2\over r^2}
 (nv) v^2\ , \label{eq:B12}
\end{equation}
and in the right side the expression
\begin{subequations}
\label{eq:B13}
\begin{equation}
 {\cal L}^{\rm N} = {8\over 15c^5} {G^3m^4\nu^2\over r^4} \biggl( 12 v^2 - 11
      (nv)^2 \biggr)\ . \label{eq:B13a}
\end{equation}
It is simple to re-write the latter expression in the well-known form
\begin{equation}
 {\cal L}^{\rm N} ={G\over 5c^5} Q^{(3)}_{ij} Q^{(3)}_{ij} + O(7) \ .
\label{eq:B13b}
\end{equation}
\end{subequations}
As this is the standard (Newtonian) quadrupole formula we
conclude that Eq.~(\ref{eq:B11}) proves the energy balance between the
loss of orbital energy $E^{\rm 2.5PN}$ of the binary and the Newtonian
energy flux carried out by the gravitational waves.  Note that it is
important that the energy which enters the left side of
the balance equation is given as an {\it instantaneous} functional of the
two world-lines of the binary (see Damour \cite{D83a} for a discussion).

\end{document}